%% file: ppc_main.tex
\documentclass[12pt]{article}

\input{preamble/preamble}
\input{preamble/math}

\input{preamble/acronyms}

\title{Holdout Predictive Checks \\ for Bayesian Model Criticism}

\usepackage{authblk}
\author[1]{Gemma E. Moran}
\author[2]{David M. Blei}
\author[3]{Rajesh Ranganath}

\affil[1]{Department of Statistics; Rutgers University}
\affil[2]{Department of Computer Science and Department of Statistics;
  Columbia University}
\affil[3]{Department of Computer Science and Center for Data Science; New York University}

\date{}                     

\begin{document}

\maketitle

\begin{abstract}
  Bayesian modeling helps applied researchers articulate assumptions
  about their data and develop models tailored for specific
  applications.  Thanks to good methods for approximate posterior
  inference, researchers can now easily build, use, and revise
  complicated Bayesian models for large and rich data.  These
  capabilities, however, bring into focus the problem of model
  criticism.  Researchers need tools to diagnose the fitness of their
  models, to understand where they fall short, and to guide their
  revision.  In this paper we develop a new method for Bayesian model
  criticism, the \gls{HPC}.  \glspl{HPC} are built on \glspl{PPC}, a
  seminal method that checks a model by assessing the posterior
  predictive distribution on the observed data.  However, \glspl{PPC}
  use the data twice---both to calculate the posterior predictive and
  to evaluate it---which can lead to uncalibrated $p$-values.  \glspl{HPC}, in contrast, compare the
  posterior predictive distribution to a draw from the population
  distribution, a heldout dataset.  This method blends Bayesian
  modeling with frequentist assessment. Unlike the \gls{PPC}, we prove
  that the \gls{HPC} is properly calibrated. Empirically, we study
  \glspl{HPC} on classical regression, a hierarchical model of
  text data, and factor analysis.
\end{abstract}

\input{sec_intro}
\input{related_work}
\input{sec_pop_pc}
\input{sec_pop_pc_theory}
\input{sec_ppc_issues}

\input{sec_experiments}

\input{sec_discussion}

\section*{Acknowledgements}
This research was supported by NSF IIS 2127869, ONR N00014-17-1-2131, ONR N00014-15-1-2209, Simons Foundation, Sloan Foundation, Open Philanthropy, NIH/NHLBI Award R01HL148248, NSF Award 1922658, NSF CAREER Award 2145542 and the Eric and Wendy Schmidt Center.

We thank Jiawei Li and Jonathan Huggins for pointing out an error in an earlier version of this manuscript.

\section*{Data Availability}
The authors confirm that the data supporting the findings of this study (except the New York Times dataset) are available within the supplementary materials. The New York Times data is available from the corresponding author, GM, upon request.

\bibliographystyle{rss}
\bibliography{bib.bib}

\appendix
\input{appendix}

\end{document}

%% file: preamble/math.tex
\newcommand{\g}{\,\vert\,}

\newcommand{\E}{\mathrm{E}}
\newcommand{\EE}[1]{\mathrm{E}\left[#1\right]}

\newcommand{\RR}{\mathbb{R}}

\newcommand{\new}{\mathrm{new}}

\DeclareMathOperator*{\argmax}{arg\,max}

\newcommand\dif{\mathop{}\!\mathrm{d}}

\newcommand{\mbw}{\bm{w}}
\newcommand{\mbx}{\bm{x}}
\newcommand{\mby}{\bm{y}}
\newcommand{\mbz}{\bm{z}}

\newcommand{\mbI}{\bm{I}}

\newcommand{\mbM}{\bm{M}}

\newcommand{\mbW}{\bm{W}}
\newcommand{\mbX}{\bm{X}}

\newcommand{\mbtheta}{\bm{\theta}}

\newcommand{\obs}{\mathrm{obs}}
\newcommand{\rep}{\mathrm{rep}}
\newcommand{\yrep}{\mby^{\rep}}

\newcommand{\ynew}{\mby^{\new}}

\newcommand{\yobs}{\mby^{\obs}}

\newcommand{\ppc}{\mathrm{ppc}}

\newcommand{\hpc}{\mathrm{hpc}}

\newcommand{\rmp}{\mathrm{p}}

\newtheorem{theorem}{Theorem}

\newtheorem{definition}{Definition}

%% file: preamble/acronyms.tex
\newacronym{LDA}{lda}{latent Dirichlet allocation}
\newacronym{PPC}{ppc}{posterior predictive check}
\newacronym{PoPC}{pop-pc}{population predictive check}
\newacronym{PoPCv1}{pop-pc-v1}{population predictive check, version 1}
\newacronym{DP}{dp}{Dirichlet process}
\newacronym{SPC}{spc}{split predictive check}
\newacronym{HPC}{hpc}{holdout predictive check}

%% file: sec_intro.tex

\glsresetall

\section{Introduction}

Thanks to good algorithms for approximate Bayesian
inference and good software for general Bayesian modeling,
statisticians can now explore and develop a large variety of Bayesian
models for a given problem.
This new ease with which we can model data has turned the
practice of Bayesian modeling into an iterative
cycle~\citep{Blei:2014,Gelman:1995,gelman2020bayesian}. More complex
models are constructed from simpler ones, and we can evaluate earlier
model ``drafts'' to guide the structure of subsequent revisions. But
this iterative process for model-building brings into sharp focus the
problem of \textit{model criticism}. How do we navigate the space of
models we can use for a problem? How do we decide when a model needs
to change?  In this paper, we develop the \gls{HPC}, a new method of
model criticism.

One of the key tools for Bayesian model criticism is the
\gls{PPC}~\citep{Guttman:1967,Rubin:1984}. In a \gls{PPC}, we locate the
observed data within the posterior predictive distribution, a
reference distribution that is determined by the model under
consideration. The spirit of a \gls{PPC} is to formalize the following
heuristic: ``If my model is good, then its posterior predictive
distribution will generate data that looks like my observations (filtered through
a diagnostic function).''

Consider an observed dataset $\yobs$ and a model with latent variables
$\mbtheta$,
\begin{align}
  \label{eq:model}
  \rmp(\mbtheta, \yobs) = \rmp(\mbtheta) \rmp(\yobs \g \mbtheta).
\end{align}
The prior is $\rmp(\mbtheta)$; the likelihood is
$\rmp(\yobs \g \mbtheta)$. The model and data combine to form the
posterior $\rmp(\mbtheta \g \yobs)$, which helps form the posterior
predictive distribution,
\begin{align}
  \label{eq:post-pred}
  \rmp(\yrep \g \yobs) = \int \rmp(\yrep \g \mbtheta) \rmp(\mbtheta \g
  \yobs) \dif \mbtheta.
\end{align}
Here the variable $\yrep$ denotes \textit{replicated data}.

To implement a \gls{PPC}, we first choose a \textit{diagnostic statistic}
$d(\mby)$, a way to measure discrepancy between a dataset $\mby$ and
the model. An example diagnostic is the average squared
distance to the posterior mean,
$d(\mby) = \frac{1}{n}\sum_{i=1}^{n} (y_i - \EE{Y \g \mby})^2$.  The
\gls{PPC} then locates $d(\yobs)$ in the \textit{reference distribution} of
$d(\yrep)$, the posterior predictive from \Cref{eq:post-pred}. One
common approach is with a posterior predictive $p$-value,
\begin{align}
  \label{eq:pppvalue}
  p_{\ppc} = \rmp(d(\yrep) \geq d(\yobs) \g \yobs).
\end{align}
\citet{Bayarri:2003} discusses other ways to measure surprise, and
\citet{Gelman:1996,Gelman:2004} discuss visual approaches to checking
the model.

But there is a crucial issue with the \gls{PPC}---it uses the data
twice. The data are first used to construct the reference distribution
of the diagnostic, the posterior predictive distribution of
\Cref{eq:post-pred}.  The data are then used again in the observed
diagnostic $d(\yobs)$, which is located within the reference, such as
with the $p$-value of \Cref{eq:pppvalue}. Essentially, a \gls{PPC}
checks how ``close'' $d(\yobs)$ is to $d(\yrep)$, a quantity that also
depends on $\yobs$.  The issue is that they can be close regardless of
whether the model is correct.

This paper introduces a solution to this ``double use of the data''
problem. The holdout predictive check (\gls{HPC}) is a new method for Bayesian model criticism,
one that combines the Bayesian \gls{PPC} with the frequentist idea of
the population distribution.  The premise of the \gls{HPC} is this:
``If my model is good, then data drawn from the posterior predictive
distribution will look like a draw from \textit{the
  population distribution}.''

Along with the data, model, and posterior predictive, consider new
data drawn from the true population distribution $\ynew \sim F$.  The
\gls{HPC} locates $d(\ynew)$ in the posterior predictive distribution
of $d(\yrep)$.  As a $p$-value, a \gls{HPC} is
\begin{align}
  \label{eq:pop-pc-pval}
  p_{\hpc} = \rmp(d(\yrep) \geq d(\ynew) \g \yobs, \ynew),
\end{align}
where $\yrep \sim \rmp(\yrep \g \yobs)$.  This quantity no longer uses
the data twice.

\begin{figure}
\begin{center}
\includegraphics[width=0.85\textwidth]{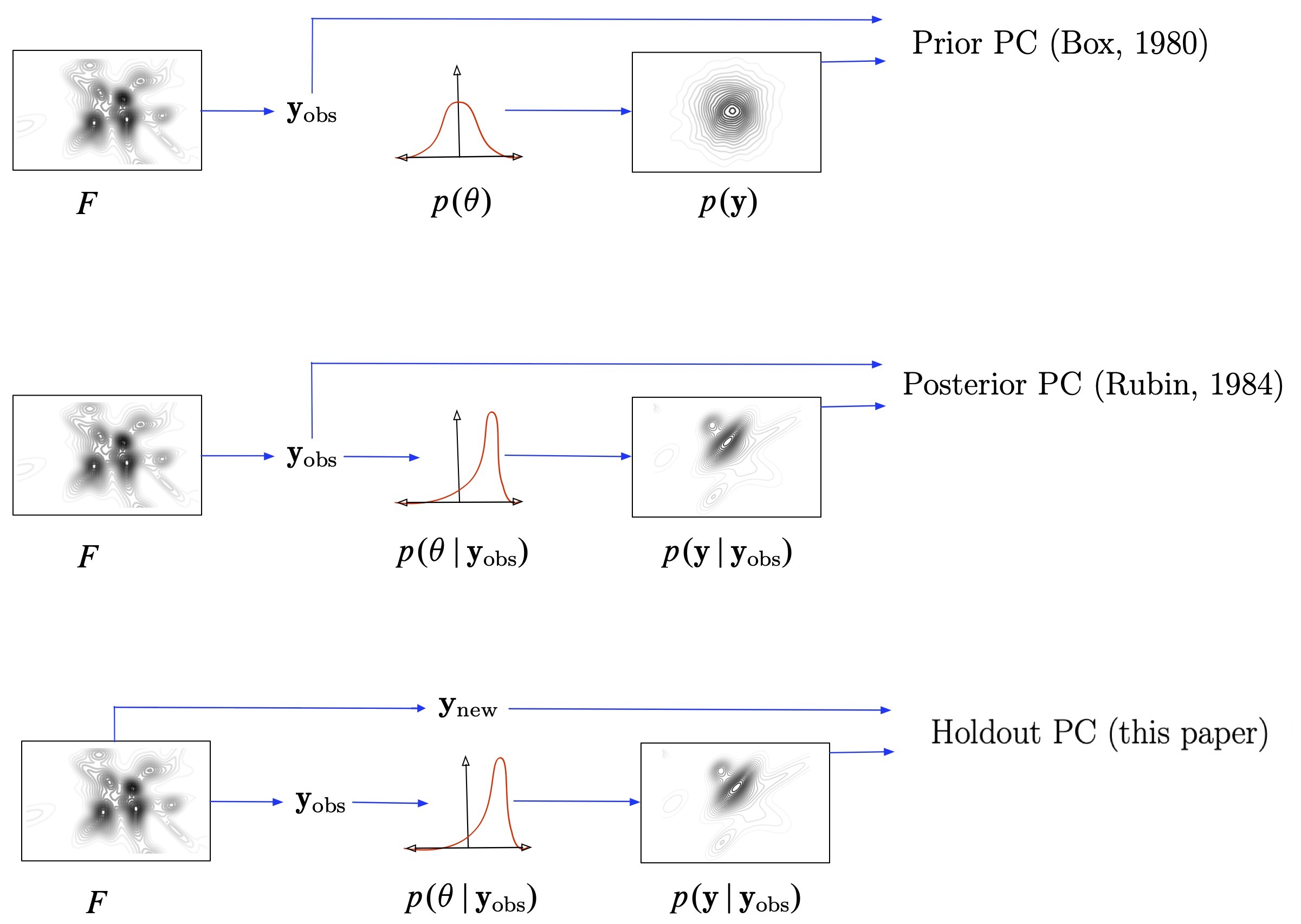}
\end{center}
\caption{A schematic diagram that relates \citet{Box:1980},
\citet{Rubin:1984}, and this paper.  This diagram posits the data come
from an unknown population distribution.  \citet{Box:1980} uses the
data as reference in the marginal distribution induced by the model
(top); \citet{Rubin:1984} uses the data as reference in the posterior
predictive distribution induced by the model (middle); our method uses
a draw from the population as a reference in the posterior predictive distribution
(bottom).}
\label{fig:methods}
\end{figure}

To implement a \gls{HPC}, we split the data into
$\mby=(\yobs, \ynew)$, assuming $\ynew$ is an independent draw from
the population distribution. We then calculate \Cref{eq:pop-pc-pval}.
As with the \gls{PPC}, the modeler can use other measures of surprise or
visual checks.  \Cref{fig:methods} displays a schematic which relates
the \gls{HPC} to both the prior predictive check \citep{Box:1980} and
the \gls{PPC} \citep{Guttman:1967,Rubin:1984}.

Why should a modeler avoid the double use of the data?  Why prefer the
\gls{HPC} of \Cref{eq:pop-pc-pval} to the \gls{PPC} of \Cref{eq:pppvalue}? We
can understand the consequences of this practice by examining some of
the frequentist properties of a predictive check.  Suppose we
repeatedly sample an observed dataset, and then check a model; note this is a frequentist situation.  We define a model to be
``correct'' when its posterior predictive distribution is equal to (or
approaches) the distribution of the observations; this is the conceit
for both a \gls{PPC} and a \gls{HPC}.

Now consider the sampling distribution of the \gls{PPC} $p$-value. There are two consequences of its double use of the data. First, this $p$-value may result in an incorrect model not being rejected; in the language of testing, it can suffer from low \textit{power}. Second, this $p$-value may not control the type I error; it may not be \textit{calibrated}. We will show below that the \gls{HPC} $p$-value does not suffer from these issues. Theoretically, under certain assumptions, we prove that it is calibrated. In the infinite data limit, and in a specific setup, the \gls{HPC} rejects an incorrect model with probability one, while calibrated versions of the \gls{PPC}~\citep{robins2000asymptotic,Hjort:2006} fail to reject the incorrect model. Empirically, we study the \gls{HPC} in several modeling situations. We find that it is better calibrated than the \gls{PPC} and that it has higher power---a \gls{HPC} more easily detects an incorrect model than either a \gls{PPC} or calibrated \glspl{PPC}~\citep{robins2000asymptotic,Hjort:2006}.

The paper is organized as follows. \Cref{sec:related} discusses the
historical development of Bayesian model evaluation and how
\glspl{HPC} fit in. \Cref{sec:pop-pcs} develops the \gls{HPC} and provides an illustrative example. \Cref{sec:calibration} proves that the \gls{HPC} $p$-value is
calibrated. \Cref{sec:problem_with_ppcs} illustrates how the ``double use of
the data'' affects the power of \glspl{PPC}, unlike the \gls{HPC}. Further, calibrating \glspl{PPC} does not resolve this power issue. 
\Cref{sec:experiments} demonstrates the \gls{HPC}
empirically on a regression model, a hierarchical document
model, and models for factor analysis. \Cref{sec:discussion} concludes the paper.


%% file: related_work.tex
\subsection{Related work}
\label{sec:related}

\glsresetall

This work builds on predictive checks, which are part of the larger
literature on Bayesian model criticism.  Predictive checks locate the
observations in a model-based distribution of data, the reference
distribution.  A brief history: Inspired by the earlier ideas of
\citet{Geisser:1975}, \citet{Box:1980} used the prior predictive
distribution as the reference. This is a prior predictive check, which
is useful for checking the conflict between prior and
likelihood~\citep{Evans:2006}.  Later, \citet{Rubin:1984} mimicked
Box's framework, but replaced the prior predictive with the posterior
predictive; this strategy is both more practical for diagnosing models
and one that is, in Rubin's language, ``Bayesianly justifiable.''
\citet{Guttman:1967} proposed the same approach.  Finally,
\citet{Gelman:1996} showed how to develop diagnostic functions of the
data---termed realized discrepancy functions---that depend on both a
data set and the latent variables.

To resolve the ``double use'' problem, researchers have proposed a
range of strategies.  One strand of research proposes to calibrate a
\gls{PPC} $p$-value post-hoc by using its empirical distribution
\citep{robins2000asymptotic,Hjort:2006}. However, calibration by
itself does not necessarily improve the power of the check, as we will show 
in \Cref{sec:problem_with_ppcs,sec:experiments}.

Alternatively, \citet{bayarri2000p} proposes the partial predictive
check. The partial predictive check calculates a predictive reference
distribution that does not depend on the diagnostic, thereby removing
the correlation between the diagnostic and the
reference. Consequently, the partial predictive check is calibrated
\citep{bayarri2000p,robins2000asymptotic}.  Finally,
\citet{johnson2007bayesian} proposed to use pivotal diagnostics to
ensure the diagnostic and reference distribution are uncorrelated.

The \gls{HPC} can be viewed as a prior predictive check where the
prior is updated based on a subset of the data. Note that this idea of
using data-dependent priors was also applied to the idea of intrinsic
Bayes factors~\citep{Berger:1996}. The intrinsic Bayes factor first
calculates a posterior using only a subset of the data. This posterior
is then treated as a prior, and the Bayes factor is calculated using
the remaining data. In a similar way, the \gls{HPC} first calculates
the posterior given the observed data, and then treats this posterior
as a prior in a prior predictive check of the new data.

The \gls{HPC} also has close connections to the partial predictive
check \citep{bayarri2000p}. Specifically, when the partial PC
diagnostic uses only a subset of the data, the partial PC can coincide
with the \gls{HPC} (see \Cref{appendix:partial-pc-discussion} for an
example).  In certain cases, a partial posterior check which computes
the diagnostic with only a subset of the data may be more efficient
than a \gls{HPC}.  This efficiency is because a partial predictive check
 may only remove a sufficient statistic of this subset of the
data, instead of the entire subset as in the \gls{HPC}. A drawback of the partial predictive check, however, is that it can be difficult to
calculate, and requires re-calculation for each diagnostic
function. Meanwhile, a \gls{HPC} is simple to implement, and the
inferred posterior can be used to check many different diagnostic
functions. A further limitation of the partial predictive check is that it can
revert to the prior predictive check when the diagnostic contains the sufficient statistics of the model (see \Cref{appendix:partial-pc-discussion} for an example).

Another closely related work is \citet{Gelfand:1992}, which develops
cross-validated checks.  A cross-validated check iteratively holds out
each data point, conditioning on the remaining data, and compares samples
from the corresponding posterior predictive distribution to the
held-out point.  Similar strategies are discussed in
\citet{Draper:1996,Marshall:2003,Larsen:2007}.  In its relation to
these other data-split checks, this paper provides a theoretical
understanding and empirical evaluation for this class of methods.

In an independent and concurrent paper,
\citet{li2022calibrated} empirically shows that a previous definition of the
\gls{HPC} (\acrshort{PoPC}, see \Cref{appendix:pop_pc_v1} for
details) is not calibrated, and proposes the \gls{SPC}. In revising
this paper, we proved the \acrshort{PoPC} is not calibrated, which
we have resolved by updating it to the \gls{HPC} in \Cref{eq:pop-pc-def}.
This \gls{HPC} is the same as the single \gls{SPC} of
\citet{li2022calibrated}, and the proof here that it is calibrated
(Theorem 1) is also similar to the proof of calibration in
\citet{li2022calibrated} for the \gls{SPC} (their Theorem 3.1(1) for
the case where the model is true). (The proofs are similar because
both build on the work of \citet{robins2000asymptotic}.)

\citet{li2022calibrated} and this paper present similar methods, but
complementary perspectives. \citet{li2022calibrated} shows the \gls{SPC} has
asymptotic power of 1 under moderate-to-major model
misspecification. It also proposes the divided \gls{SPC}, which
considers \gls{SPC} $p$-values for multiple different splits of the
data. In this paper, we examine the ``double use of the data'' problem
of the \gls{PPC} in greater detail. Specifically, we illustrate that
post-hoc empirical calibration procedures for the \gls{PPC}, while producing uniform $p$-values under the null hypothesis, do not resolve the ``double use of the data''
problem in terms of detecting model misspecification.  Finally, we
consider different diagnostics from \citet{li2022calibrated} - the
$\chi^2$ diagnostic and latent Dirichlet allocation log-likelihood -
and we provide empirical evidence that they are calibrated; these
diagnostics are not covered by the calibration theory for the
\gls{HPC}, nor the \gls{SPC}.

Finally, the \gls{HPC} relates to metrics which assess generalization error,
though it is also distinct from these ideas. Assuming a stationary
data generating process, the \gls{HPC} asks: does my model produce data that
“looks like” future independent draws from this process? This binary
outcome is in contrast to continuous metrics, such as the widely
applicable information criterion \citep[WAIC,][]{watanabe2009algebraic,watanabe2010asymptotic} or the
deviance information criterion \citep[DIC,][]{spiegelhalter2002bayesian}. While the WAIC and DIC can be used to select from multiple models, they do not assess the adequacy of a single model, which is of
interest in this work.


%% file: sec_pop_pc.tex
\glsresetall

\section{Holdout Predictive Checks}
\label{sec:pop-pcs}

The \gls{HPC} checks a Bayesian model by considering a true population
distribution: ``If my model is good then data
drawn from the posterior predictive distribution will look like a draw
from the true population  (filtered through a diagnostic function).''

The ingredients of a \gls{HPC} are observed data $\yobs$, replicated
data $\yrep$ from the posterior predictive distribution
(\Cref{eq:post-pred}), and new data $\ynew$, drawn from the true
population distribution $F$.  As for a \gls{PPC},
each check involves a diagnostic statistic $d(\mby)$, which measures
misfit between $\mby$ and the model.  The \gls{HPC} uses  new data
$\ynew$ to check if a draw from the population $F$ is close to the posterior predictive distribution
$\rmp(\yrep \g \yobs)$, in terms of the
diagnostic.  If so, then the posterior predictive captures the data
well, and the model passes the check.

\begin{definition}[Holdout predictive check, \gls{HPC}] \label{def:pop-pc}
Consider observed data $\yobs$, its posterior predictive distribution $\rmp(\yrep \g \yobs)$, and a diagnostic statistic $d(\mby)$. Suppose we have $\ynew$ drawn from the population distribution of the data. As a $p$-value, the holdout predictive check is:
\begin{align}
  p_{\hpc} = \rmp(d(\yrep) \geq d(\ynew) \g \yobs, \ynew), \label{eq:pop-pc-def}
\end{align}
where $\yrep \sim \rmp(\yrep \g \yobs)$.
\end{definition}

To implement a \gls{HPC}, we split the data into $\mby=(\yobs, \ynew)$ and calculate \Cref{eq:pop-pc-def}.

The diagnostic is a function of the data that measures
model misfit.  For
example, one diagnostic is the conditional negative log-likelihood,
\begin{align}
  d(\mby) \triangleq -\frac{1}{n} \sum_{i=1}^n \log \rmp(y_i \g \yobs).
\end{align}
In the context of a \gls{PPC}, this diagnostic is discussed in
\citet{Lewis:1996}.

\begin{algorithm}
\caption{Holdout predictive check} \label{alg:pop-pc}
\KwInput{data $\mby=\{\yobs, \ynew\}$, diagnostic $d(\cdot)$, \# replicates $R$ }
\KwOutput{holdout predictive check $p$-value}
\For{$r = 1,\dots, R$}{
draw samples from the posterior $\theta_r \sim \rmp(\theta|\yobs)$\;
draw posterior predictive data $\mby_r^{\mathrm{rep}}\sim\rmp(\yrep|\theta)$\;
}
\BlankLine
compute the empirical \gls{HPC} $p$-value
   \begin{align*}
   p_{\hpc} &= \frac{1}{R}\sum_{r=1}^R 1\left[d(\mby_r^{\mathrm{rep}}, \theta_r) > d(\ynew, \theta_r)\right];
   \end{align*}
\KwRet{$p_{\hpc}$}
\end{algorithm}

\cite{Meng:1994,Gelman:1996} discuss realized diagnostics $d(\mby,\theta)$, those that
also depend on the latent variables.  A realized diagnostic measures
the strength of the connection between latent variables and a data
set.  An example is the negative log-likelihood
\begin{align}
d(\mby, \theta) \triangleq - \log \rmp(\mby \g \theta).
\end{align}

Consider a joint distribution of latent variables and data,
\begin{align}
  \rmp(\theta, \mby) = \rmp(\theta) \rmp(\mby \g \theta)
\end{align}
and a realized diagnostic $d(\theta, \mby)$.  The
\gls{HPC} is
\begin{align}
  p_{\hpc} = \rmp(d(\theta, \yrep) > d(\theta, \ynew)
  \g \yobs, \ynew).
\end{align}
This probability is under a distribution that draws the latent
variable from the posterior and the replicated data from the likelihood
given the latent variable, 
\begin{align}
  \rmp(\theta, \yrep \g \yobs) =
  \rmp(\theta \g \yobs) \rmp(\yrep \g \theta).
\end{align}
The \gls{HPC} procedure is detailed in \Cref{alg:pop-pc}.

\subsection{Example} \label{sec:pop-pc-regression-intro}

To illustrate the \gls{HPC}, we now apply a ridge regression model to synthetic data. The model is: 
\begin{align}
\bm{\theta} &\sim \textrm{Normal}(0, c\bm{I}_p), \\
\sigma^2 &\sim \textrm{Inverse-Gamma}(1/2, 1/2), \\
y_i |\bm{\theta}, \sigma^2&\stackrel{iid}{\sim} \textrm{Normal}(\bm{\theta}^\top \bm{x}_i, \sigma^2), \quad i = 1,\dots, n. \label{eq:ridge_model}
\end{align}

We take $n = 50$ and $p=100$.  The covariates, $\bm{x}_i$, are drawn as uniform random variables on $[0,1]$.  Meanwhile, the true coefficients, $\theta$, have five entries equal to 3.5 and the remaining 95 entries drawn from a standard normal distribution. Consequently, the coefficients $\theta$ have many entries close to zero, and a few large entries.  The true $\sigma^2=1$ (note that $\sigma^2$ is treated as unknown during inference, however).

What is the impact of the prior variance parameter $c$ on the adequacy of the model?  If $c$ is too large, the estimated coefficients will overfit to the data and not generalize well on new data. Here, we show that a \gls{HPC} can detect this poor model fit while a \gls{PPC} cannot. Note that we are not advocating to use the \gls{HPC} to choose $c$; we are merely emphasizing that in the well-studied setting of ridge-regression, the \gls{HPC} can detect poor model fit while the \gls{PPC} cannot.

To check the model, we use the $\chi^2$ diagnostic function:
\begin{align}
\label{eq:chi-squared} d(\bm{y}, \bm{\theta}, \sigma^2) = \frac{1}{N}\sum_{i=1}^N
\frac{ (y_i - \E[y_i | \bm{\theta}, \sigma^2])^2}{\text{Var}(y_i | \bm{\theta}, \sigma^2)},
\end{align}
which is a sum of standardized residuals.  We conduct posterior inference using a Gibbs sampler. 

\begin{figure}[t]
\centering
\includegraphics[width = 0.7\textwidth]{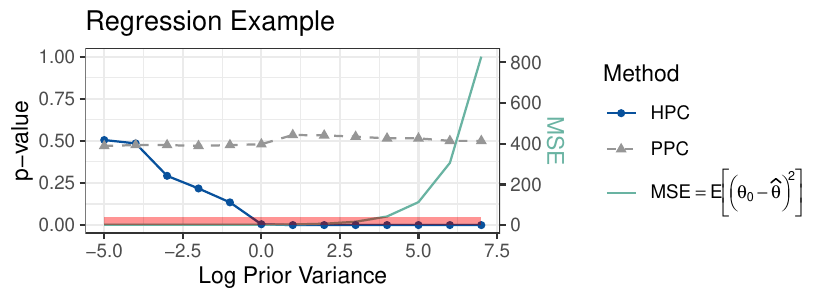}
\caption{When the log prior variance $c$ is small, both the \gls{HPC} and \gls{PPC} correctly retain the model; when $c$ is large, the \gls{HPC} correctly rejects the model, while the \gls{PPC} incorrectly retains the model. } \label{fig:pop_pc_example}
\end{figure}

\Cref{fig:pop_pc_example} shows the \gls{HPC} and \gls{PPC} $p$-values for different values of $c$. Also plotted is the mean squared error:
\begin{align}
\text{MSE} = \mathbb{E}\left[\left(\theta_0 - \theta\right)^2\right]
\end{align}
where the expectation is with respect to the posterior $\rmp(\theta|\mby, \mbX)$ and $\theta_0$ is the true value of the coefficents.

When $c$ is small, there is more regularization of the coefficient estimates. This regularization prevents overfitting and results in small mean squared error.  For these small values of $c$, both the \gls{PPC} and \gls{HPC} both correctly retain the model. When $c$ is large, however, there is less regularization of the coefficients.  Consequently, the model overfits to the data and the MSE is large.  For these large values of $c$, the \gls{HPC} correctly rejects the model. The \gls{PPC}, however, does not reject the model; the \gls{PPC} does not detect overfitting, unlike the \gls{HPC}.

%% file: sec_pop_pc_theory.tex
\glsresetall

\section{The asymptotic distribution of \textsc{HPC} $p$-values} \label{sec:calibration}

In this section, we study the \gls{HPC} $p$-value asymptotic sampling distribution. If a $p$-value is uniformly distributed when the model is correct, the $p$-value is said to be \emph{calibrated}.  For a class of diagnostic functions, we prove that the \gls{HPC} $p$-value is calibrated (\Cref{theorem:pop_pc_ideal_calibration}). 

Calibration is a frequentist property, not a Bayesian one. Although the \gls{HPC} checks Bayesian models, it is still important to determine if its $p$-values are calibrated. Calibration helps us interpret a $p$-value: If the distribution of $p$-values is uniform, a $p$-value of 0.4 would not be surprising, while if the distribution was concentrated around 0.5, the value 0.4 would be surprising.  For a calibrated check, if we decide to reject a model when its  $p$-value is less than $\alpha$, then the correct model will only fail
such a check with that probability.

Before stating our \gls{HPC} calibration result, we introduce some 
notation. The data are $\mby = (y_1,\dots, y_n)$ where $y_i$ are mutually independent random variables with density $f(y; \theta)$ where $\theta\in \Theta\subset \mathbb{R}^p$. 

We prove the \gls{HPC} is calibrated for the same class of diagnostic functions $d(\mby)$ that are considered by \citet{robins2000asymptotic}.  Specifically, we consider diagnostic functions $d(\mby)$ that are asymptotically normal with asymptotic mean $\nu(\theta)$ and asymptotic variance $\sigma^2(\theta)$ when the model is correct (i.e. the density of $\mby$ is $f(\mby;\theta)$):
\begin{align}
n^{1/2}\left[\frac{d(\mby) - \nu(\theta)}{\sigma(\theta)}\right]\leadsto N(0,1),\label{eq:pop_pc_diagnostic_convergence}
\end{align}
where $\leadsto$ denotes convergence in distribution. 

\Cref{theorem:pop_pc_ideal_calibration} proves that the \gls{HPC} $p$-values are asymptotically uniform when the model is correct. Unlike the \gls{PPC}, the \gls{HPC} $p$-values are calibrated.  Note our result relies on standard regularity conditions that are detailed in \Cref{appendix:calibration_proof}.

\begin{theorem}\label{theorem:pop_pc_ideal_calibration}
Assume \Cref{eq:pop_pc_diagnostic_convergence} holds and assume the regularity conditions detailed in \Cref{appendix:calibration_proof}. Under the distribution $f(\mby; \theta_0)$, the \gls{HPC} $p$-value can be written as:
\begin{align}
p_{\hpc}(\mby) &= 1 - \Phi(Q)+o_P(1), 
\end{align}
where $o_P(1)$ denotes a random variable converging to zero in probability, $\Phi$ is the standard normal cdf, and $Q\sim N(0,1)$. Consequently, the \gls{HPC} $p$-values are calibrated.
\end{theorem}
The proof of \Cref{theorem:pop_pc_ideal_calibration} is in \Cref{appendix:calibration_proof}.

\Cref{theorem:pop_pc_ideal_calibration} proves that \gls{HPC} $p$-values are calibrated for realized diagnostics that are asymptotically normal. In experiments (\Cref{sec:experiments}) we consider diagnostics that are not asymptotically normal, including the $\chi^2$ diagnostic and the log-likelihood of latent Dirichlet allocation \citep{Blei:2003b}.  With these diagnostics, we show empirically the \gls{HPC} still has good calibration properties.

\subsection{Data splitting for the \textsc{HPC}}

We acknowledge that splitting the data may reduce power to detect model misspecification. This reduction in power may be avoided by using a {PPC} with a diagnostic that is independent of the model parameters.  However, it can be difficult to verify such independence, and so the {HPC} allows practitioners more flexibility in diagnostic choice.

A separate aspect of data splitting is that the value of $p_{\mathrm{hpc}}$ will depend on the particular split of the data. However, Theorem 1 provides an asymptotic guarantee that $p_{\mathrm{hpc}}$ is uniformly distributed over any split of the data. Further, in finite sample settings, our experiments show that the HPC is approximately uniform.

Finally, we note that the \gls{HPC} data splitting requirement limits us to diagnostic functions that can be calculated on a split of the data--we are unable to use diagnostics that depend jointly on all observations.


%% file: sec_ppc_issues.tex
\glsresetall

\section{A comparison of predictive checks}\label{sec:problem_with_ppcs}

In the previous section, we studied the \gls{HPC} $p$-value distribution when the model is correct. In this section, we consider a specific model and diagnostic function to illustrate the properties of the \gls{HPC} $p$-value when the model is incorrect. Under this setup, we show the \gls{HPC} has high power--it detects this model misfit with probability one in the infinite data limit, while the \gls{PPC} and calibrated versions of the \gls{PPC}~\citep{robins2000asymptotic,Hjort:2006} cannot detect this model misfit; that is, calibration by itself cannot improve the power of a test. 

The setup is: we have data $\yobs = \{y_i\}_{i=1}^n$ for which we posit a Gaussian model:
\begin{align}
y_i  \sim N(\mu, \sigma^2), \quad i=1,\dots,n \label{eq:h0_mean}
\end{align}
with unknown mean $\mu\in\mathbb{R}$ and known $\sigma^2$.

We want to check the suitability of this Gaussian model with a Bayesian predictive check. 
Unlike in a frequentist hypothesis test, the parameters of the model are not set to pre-specified values. That is, in \Cref{eq:h0_mean} we are not checking a specific value of the mean $\mu$, but the appropriateness of the Gaussian model. 

In this section, we consider a specific alternative data distribution that would not fall in the Gaussian model (\Cref{eq:h0_mean}), so we can explicitly compare the power of predictive checks. This data distribution is a Cauchy distribution with location $x_0\in\mathbb{R}$ and scale $\gamma=1$:
\begin{align}
y_i \sim \text{Cauchy}(x_0, 1), \quad i=1,\dots,n. \label{eq:h1_mean}
\end{align}

Specifically, we analyze how the \gls{PPC}, calibrated \glspl{PPC} and the \gls{HPC} perform when checking the Gaussian model (\Cref{eq:h0_mean}) if the underlying data is actually Cauchy (\Cref{eq:h1_mean}).

We use a mean diagnostic function: $d(\yobs) = \overline{\mby}^{\textrm{obs}}$.  We chose this diagnostic function as it results in an analytic form for the \gls{PPC}. This analytic form allows us to demonstrate clearly how the \gls{PPC} fails to detect model misfit. While there are other choices of diagnostic function for which the \gls{PPC} will detect model misfit in this scenario (e.g. the maximum), our main point is that the \gls{HPC} will avoid the failure modes of the \gls{PPC} no matter the choice of diagnostic function, allowing for more flexibility in diagnotic choice for practictioners.\footnote{Others have also used the mean diagnostic to check a normal model (see, for example, the `8 Schools' example in Section 6.5, \citet{gelman2013bayesian}).}

\subsection{Posterior predictive checks} 

We first consider the distribution of the \gls{PPC} $p$-value with the mean diagnostic which checks the Gaussian model in \Cref{eq:h0_mean}.  We show that this particular \gls{PPC} $p$-value is degenerate at 0.5 regardless of whether the data is Gaussian or Cauchy.  This degeneracy of the \gls{PPC} $p$-value is problematic for both calibration and power. Specifically, the \gls{PPC} $p$-value is not uniform and thus not calibrated. Further, the \gls{PPC} has low power - it cannot detect model misfit when the data is actually Cauchy.

For a prior on $\mu$, we take $\mu\sim N(\mu_0, \sigma_0^2)$ with fixed hyperparameters $\mu_0\in\mathbb{R}, \sigma_0^2\in\mathbb{R}^+$. 

Concretely, the posterior predictive $p$-value with diagnostic $d(\yobs) = \overline{\mby}^{\textrm{obs}}$ is:
\begin{align}
p_{\ppc}= \rmp(d(\yrep) > d(\yobs) | \yobs) &= 1 - \Phi\left(\frac{\overline{\mby}^{\obs} - \rho_n \overline{\mby}^{\obs}- (1 - \rho_n)\mu_0}{\sqrt{(1 + \rho_n)\sigma^2/n}} \right), \label{eq:ppc_mean}
\end{align}
where $\rho_n = \sigma_0^2/(\sigma_0^2 + \sigma^2/n)$ (for details, see \Cref{appendix:problem_with_ppcs}).  

Ideally, the \gls{PPC} would check some aspect of model misfit. However, the \gls{PPC} in \Cref{eq:ppc_mean} is only checking how ``close'' the posterior mean $\mathbb{E}[\mu|\yobs]=\rho_n \overline{\mby}^{\obs}+ (1 - \rho_n)\mu_0$ is to the MLE $\overline{\mby}^{\obs}$.  Whether the posterior mean is close to the MLE is a property of the model, unrelated to the fit of the model to the data. That is, the posterior mean can be close to the MLE, whether or not the underlying data is actually Gaussian.  

To further illustrate this problem with the \gls{PPC}, consider the asymptotic distribution of $p_{\ppc}$. In \Cref{eq:ppc_mean}, the numerator is $\overline{\mby}^\obs - \rho_n \overline{\mby}^\obs- (1 - \rho_n)\mu_0 = O(1/n)$ and the denominator is $O(1/\sqrt{n})$. Consequently, the integrand goes to zero as $O(1/\sqrt{n})$ and the $p_{\ppc}$ converges to 0.5. What is important to note is that $p_{\ppc}$ converges to 0.5 regardless of whether the data is Gaussian or Cauchy.

\subsection{Empirically calibrated  \textsc{PPC}s can have low power}

To fix the calibration of \gls{PPC} $p$-values, a number of post-processing strategies have been proposed \citep{robins2000asymptotic,Hjort:2006}. In this section, we show that such post-hoc calibration techniques do not improve the power of the \gls{PPC} to detect model misfit. 

Essentially, these calibration techniques cannot detect model misfit because for any random variable, we can calibrate it by using its cdf to transform it to a uniform distribution. If the original random variable does not detect model misfit, this transformation will not provide additional power to detect model misfit.

We make the above point more concrete by again considering \Cref{eq:h0_mean} and analyzing the following two post-hoc calibration methods:
\begin{itemize}
    \item \citet{robins2000asymptotic} proposed to locate the observed \gls{PPC} $p$-value in the empirical distribution of the \gls{PPC} $p$-values. The empirical distribution of the \gls{PPC} $p$-values is calculated by treating draws from the posterior predictive as replicates from the true model and then calculating their \gls{PPC} $p$-values.
    \item \citet{Hjort:2006} also propose to locate the observed \gls{PPC} $p$-value in an empirical reference distribution. Unlike \citet{robins2000asymptotic}, however, \citet{Hjort:2006} use draws from the prior predictive distribution to calculate an empirical reference distribution (instead of draws from the posterior predictive).  
\end{itemize}

First, we show that for the mean diagnostic, the \citet{robins2000asymptotic} empirically calibrated \gls{PPC} $p$-value has the same distribution regardless of whether the data is Gaussian or Cauchy. Consequently, the check cannot detect model misfit.

The \citet{robins2000asymptotic} empirically calibrated \gls{PPC} $p$-value is
\begin{align}
\rmp\left(\text{ppc}(\yrep) > \text{ppc}(\yobs) | \yobs \right).
 \end{align}

To compute this empirically calibrated $p$-value, we first need to determine the empirical distribution of $\text{ppc}(\yrep)$--these are the \gls{PPC} values when $\mby^{\mathrm{rep}}$ is treated as an observed dataset.  We use the notation $\mby^{\mathrm{rep,rep}}$ to denote a draw from the posterior predictive distribution conditioned on $\mby^{\mathrm{rep}}$; that is, the posterior predictive where $\mby^{\mathrm{rep}}$ is ``observed data''. More specifically, $\text{ppc}(\yrep)$ is calculated as
 \begin{align}
\text{ppc}(\yrep) = \frac{1}{R}\sum_{r=1}^R \mathbbm{1}(\overline{\mby}^{\mathrm{rep,rep}}_r > \overline{\mby}^\rep), \quad {\mby}^{\mathrm{rep,rep}}_r \sim \rmp(\mby^{\mathrm{rep,rep}}|\yrep),\label{eq:ppc_empirical}
 \end{align}
with $R$ draws from the posterior predictive distribution. Now, suppose $\rho_n\to1$. Then, $\sqrt{n}(\overline{\mby}^{\mathrm{rep,rep}}-\overline{\mby}^\rep)|\overline{\mby}^\rep\sim N(0, 2\sigma^2)$.  In this case, the probability that $\overline{\mby}^{\textrm{rep,rep}}$ is greater than its mean is 0.5 and so the sum of indicators is a binomial random variable:
 \begin{align}
\sum_{r=1}^R \mathbbm{1}(\overline{\mby}_r^{\mathrm{rep,rep}} > \overline{\mby}^\rep) \sim \text{Binomial}(R,0.5). \label{eq:mean_ppc_rep_rep}
 \end{align}
With the normal approximation to the binomial distribution,
 \begin{align}
\rmp\left(\text{ppc}(\yrep) > \text{ppc}(\yobs) | \yobs \right)=1-\Phi\left(2\sqrt{R}\left[\text{ppc}(\yobs) - 0.5\right]\right).
 \end{align}
 To determine the distribution of the empirically calibrated \gls{PPC}, consider the distribution of $\text{ppc}(\yobs)$. Similarly to \Cref{eq:mean_ppc_rep_rep}, for large $R$,
 \begin{align}
2\sqrt{R}[\text{ppc}(\yobs) - 0.5] \sim N(0, 1),
 \end{align}
 again using the normal approximation to the binomial distribution. Then, the empirically calibrated \gls{PPC} is uniform:
\begin{align}
\rmp\left(\text{ppc}(\yrep) > \text{ppc}(\yobs) | \yobs \right)\sim U[0,1].
\end{align}

The key point is that the calibrated \gls{PPC} is uniform regardless of whether the data is Gaussian or Cauchy. This is because for both data distributions, we have $\sqrt{n}(\overline{\mby}^{\mathrm{rep,rep}}-\overline{\mby}^\rep)|\overline{\mby}^\rep\sim N(0, 2\sigma^2)$ and $\sqrt{n}(\overline{\mby}^\rep- \overline{\mby}^\obs)|\overline{\mby}^\obs\sim N(0, 2\sigma^2)$. In other words, here, the empirically calibrated \gls{PPC} does not depend on the underlying distribution of $\overline{\mby}^\obs$ and so it cannot detect model misfit.

We now show that \citet{Hjort:2006}'s calibrated $p$-value (cppp) also has the same distribution regardless of whether the data is Gaussian or Cauchy. Consequently, the check cannot detect model misfit.

The cppp is:
\begin{align}
\rmp(\textrm{ppc}(\mby) > \textrm{ppc}(\yobs)|\yobs), \quad \text{where} \quad\mby\sim \rmp(\mby;\theta),\ \theta \sim \rmp(\theta),
\end{align}
and $\textrm{ppc}(\cdot)$ is defined in \Cref{eq:ppc_empirical}.

Similarly to \Cref{eq:mean_ppc_rep_rep}, 
\begin{align}
R\cdot\textrm{ppc}(\mby) = \sum_{r=1}^R \mathbbm{1}(\overline{\mby}_r^\rep > \overline{\mby})\sim\text{Binomial}(R,0.5),
\end{align}
where $\mby^\rep_r \sim \rmp(\mby^\rep|\mby)$. 

Following a similar argument to the empirically calibrated \gls{PPC} of \citet{robins2000asymptotic}, 
\begin{align}
\rmp(\textrm{ppc}(\mby) > \textrm{ppc}(\yobs)|\yobs)\sim U[0,1].
\end{align}
Again, this will hold for regardless of whether the data is Gaussian or Cauchy and so the cppp will fail to detect model misfit. 

In \Cref{sec:experiments}, we consider a linear regression example and demonstrate that these post-processing techniques yield calibrated \gls{PPC} $p$-values, but that these $p$-values fail to detect model misspecification. In contrast, the \gls{HPC} is calibrated and does detect model misspecification.

\subsection{Comparison with holdout predictive checks}

We have reviewed the ``double use of the data'' problem with \gls{PPC}s, which can result in both uncalibrated $p$-values and minimal power to detect model misfit.  These $p$-values can be empirically calibrated, but the calibration procedures do not necessarily improve the power of the test. In contrast, the \gls{HPC} can detect model misspecification in situations where a \gls{PPC} or calibrated \gls{PPC} cannot. 

Consider the example in \Cref{eq:h0_mean}. When the data is actually Gaussian, the \gls{HPC} $p$-value is:
\begin{align} 
\rmp(d(\yrep) > d(\ynew)|\yobs, \ynew) &= 1 - \Phi\left(\frac{\overline{\mby}^\new - \rho_n\overline{\mby}^\obs - (1-\rho_n)\mu_0}{\sqrt{(1 + \rho_n)\sigma^2/n}}\right). \label{eq:mean_pop_pc}
\end{align}
As $n\to\infty$, we have 
\begin{align}
\sqrt{n}[\overline{\mby}^\new - \rho_n\overline{\mby}^\obs - (1-\rho_n)\mu_0] \sim N(0, 2\sigma^2).
\end{align}
Then, 
\begin{align}
\rmp(d(\yrep) > d(\ynew)|\yobs, \ynew) = 1 - \Phi(Z),
\end{align}
where $Z\sim N(0,1)$, and so the \gls{HPC} $p$-value is uniform and calibrated. (This is an example of the more general calibration result proved in \Cref{theorem:pop_pc_ideal_calibration}).

Now suppose the data is actually Cauchy (\Cref{eq:h1_mean}). Then, we have $\overline{\mby}^\new - \overline{\mby}^\obs \sim \text{Cauchy}(0, 2)$.  
We prove that the \gls{HPC} has asymptotic power of one--that is, \gls{HPC} will reject the Gaussian model if the data is actually Cauchy with probability one. Let $z_{\alpha}$ denote the $\alpha$-quantile of a standard Gaussian distribution. Then for the two-sided \gls{HPC}, the power at rejection level $\alpha$ is:
\begin{align}
\text{Power} &= \rmp\left(\Phi\left(\frac{\overline{\mby}^\new - \overline{\mby}^\obs}{\sqrt{2\sigma^2/n}}\right) \leq \alpha/2\right) + \rmp\left(\Phi\left(\frac{\overline{\mby}^\new - \overline{\mby}^\obs}{\sqrt{2\sigma^2/n}}\right) \geq 1-\alpha/2\right) \\
&= \rmp\left(\overline{\mby}^\new - \overline{\mby}^\obs \geq z_{1-\alpha/2} \sqrt{\frac{2\sigma^2}{n}}  \right) 
+
\rmp\left(\overline{\mby}^\new - \overline{\mby}^\obs \leq z_{\alpha/2} \sqrt{\frac{2\sigma^2}{n}}\right) \\
&\stackrel{n\to\infty}{\to} 1.
\end{align}

\begin{figure}
\begin{subfigure}[b]{\textwidth}{}
\centering
\includegraphics[width=0.8\textwidth]{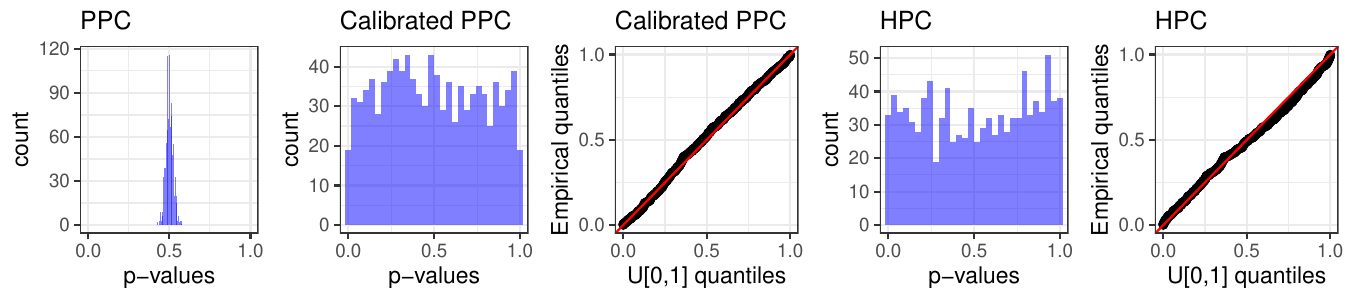}
\caption{Distribution of $p$-values when the data is Gaussian (\Cref{eq:h0_mean}).}
\end{subfigure}
\begin{subfigure}[b]{\textwidth}
\centering
\includegraphics[width=0.8\textwidth]{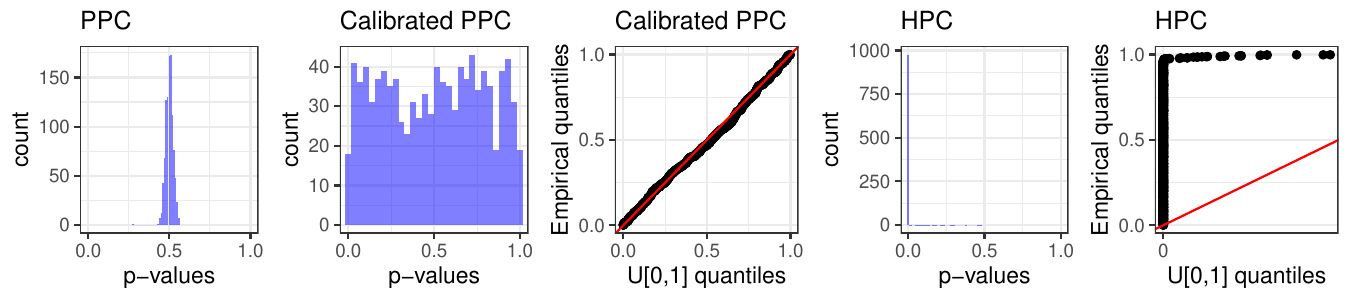}
\caption{Distribution of $p$-values when the data is Cauchy (\Cref{eq:h1_mean}).}
\end{subfigure}
\caption{In the Gaussian mean example, the \gls{HPC} detects model misspecification (i.e. when the data is actually Cauchy) while the \gls{PPC} and calibrated \gls{PPC} do not.}\label{fig:mean_example}
\end{figure}

To further illustrate these points, the empirical distributions of the \gls{PPC}, calibrated \gls{PPC} and \gls{HPC} for the simple mean example are displayed in \Cref{fig:mean_example}. We see the \gls{PPC} is concentrated around 0.5 both when the data is Gaussian and when the data is Cauchy. Moreover, the calibrated \gls{PPC} is uniform for both Gaussian- and Cauchy-distributed data. In contrast, the \gls{HPC} is uniform when the data is Gaussian, and concentrated around 0 when the data is Cauchy. That is, the \gls{HPC} detects model misspecification while the \gls{PPC} or calibrated \gls{PPC} do not.

\subsection{Choosing diagnostic functions}

We showed that the \gls{PPC} for checking a Gaussian model with a mean diagnostic is degenerate, even under the alternative hypothesis when the true data distribution is Cauchy. 

In general, when does this degeneracy occur for \gls{PPC} $p$-values? The \gls{PPC} becomes degenerate when the diagnostic is perfectly correlated with the model parameters \citep{robins2000asymptotic}. More specifically, let $\theta(\yobs)$ be the posterior mean of the model parameters, conditioned on the observed data. Then under the conditions of \Cref{theorem:pop_pc_ideal_calibration}:
\begin{itemize}
\item if the diagnostic $d(\yobs)$ is perfectly correlated with $\theta(\yobs)$, then the \gls{PPC} will converge to 0.5 under both the null and alternative hypothesis;
\item if the diagnostic is independent of $\theta(\yobs)$, the \gls{PPC} $p$-values are uniformly distributed under the null hypothesis and thus calibrated \citep{robins2000asymptotic}.  
\end{itemize}
However, it is often difficult to determine the degree of association between $\theta(\yobs)$ and $d(\yobs)$, and thus difficult to assess whether the \gls{PPC} will detect model misfit.  In contrast, we proved that the \gls{HPC} is calibrated for all normally-distributed diagnostic functions (\Cref{theorem:pop_pc_ideal_calibration}). This result allows the \gls{HPC} much more flexibility in the choice of diagnostic than the \gls{PPC}.

The choice of diagnostic function also has implications for the power of the test. For the Gaussian/Cauchy example studied in this section, the mean diagnostic can detect model misfit because under $H_0$, $\overline{\ynew}-\overline{\yobs}\sim N(0,2)$, while under $H_1$, $\overline{\ynew}-\overline{\yobs}\sim \text{Cauchy}(0,2)$--interestingly, it is the variance in the mean diagnostic that allows us to detect model misfit.

Other choices of diagnostic functions will also change the power of the test. As a trivial example, if we had chosen $d(\mby)=c$ for some constant $c$, the distribution of $d(\mby)$ would be the same under both $H_0$ and $H_1$ and so we cannot detect model misfit.

As a final example of how the diagnotic function may affect power, suppose we take $d(\mby)=\max \mby$. When will we have the same distribution of the test-statistic under both $H_0$ and $H_1$? This may occur if the models in $H_0$ and $H_1$ are such that their maxima converge to the same distribution. This is possible given the extreme value theorem, which tells us that for a variety of distributions, the distribution of their maxima is either Gumbel, Fr\'{e}chet or Weibull.

%% file: sec_experiments.tex
\glsresetall

\section{Empirical Study}
\label{sec:experiments}

We study~\acrlongpl{HPC} on a regression model, a hierarchical
model of documents, and a factor model.

\begin{itemize}
\item In the regression model study:
\begin{enumerate}[label=(i)]
	\item We show empirically that the \gls{HPC} detects overfitting when there is insuficient regularization. In contrast, the \gls{PPC}, as well as the calibration suggestions of \citet{robins2000asymptotic} and \citet{Hjort:2006}, do not detect overfitting when there is insufficient regularization. 
	\item We show empirically that the \gls{HPC} $p$-values are approximately uniform when the model has sufficient regularization (i.e. the \gls{HPC} is calibrated). In contrast, the \gls{PPC} $p$-values are not calibrated. 
\end{enumerate}
\item For the hierarchical model of documents:
\begin{enumerate}[label=(i)]
	\item On synthetic data, we show empirically that the \gls{HPC} $p$-values are approximately uniform when the data actually comes from the model (i.e. the \gls{HPC} is calibrated).
	\item On a collection of documents from the \emph{New York Times}, we show that the \gls{HPC} detects model misfit due to overfitting, while the \gls{PPC} does not.
\end{enumerate}
\item In the factor model study:
\begin{enumerate}[label=(i)]
	\item When the model is correct, we show empirically that the \gls{HPC} $p$-values show much greater variability than the \gls{PPC} $p$-values, which are centered around 0.5.
	\item When the model is incorrect, we show that the \gls{HPC} correctly reject the model at a much higher rate than the \gls{PPC}.
\end{enumerate}
\end{itemize}

\subsection{Bayesian Ridge Regression} \label{sec:ridge_regression}

In this section, we return to the regression example in \Cref{sec:pop-pc-regression-intro}.  We now compare the \gls{HPC} with a variety of methods for Bayesian model criticism for a range of different values of the prior variance $c$.  Again, we set $n=50$ and $p=100$.

In the large $c$ scenario, we expect that a posterior predictive check will not be able to detect the overfitting of the data. Moreover, as discussed in \Cref{sec:problem_with_ppcs}, we expect the calibration strategies of \citet{robins2000asymptotic} and \citet{Hjort:2006} will also not detect this overfitting. 
In contrast, we expect that a holdout predictive check will be able to detect overfitting, and reject the model with large prior variance values.  Recall that we are not using the \gls{HPC} to choose the value of the regularization parameter $c$; we are simply emphasizing that when there is clear overfitting, the \gls{HPC} is able to detect this poor model fit, while the \gls{PPC} cannot.

We also consider prior predictive checks \citep{Box:1980}.

\begin{figure}
\centering
	\includegraphics[width=\textwidth]{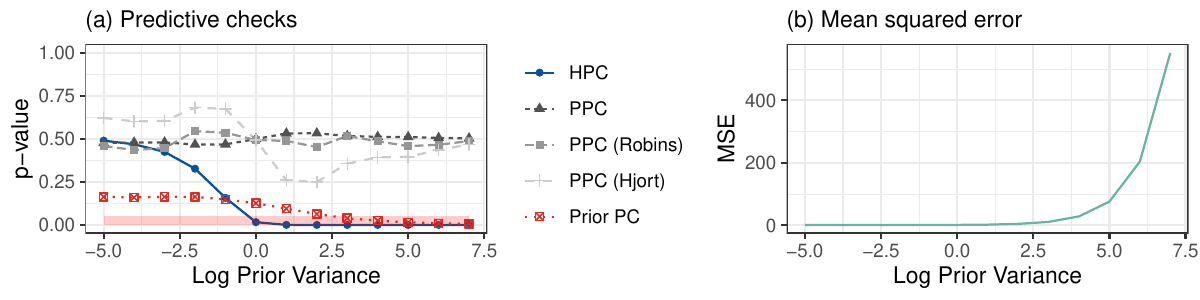}
	\caption{Bayesian ridge regression: \gls{HPC} detects overfitting when there is little regularization (large $c$), while the \gls{PPC} (including calibrated versions) does not.  (a) average $p$-values over different $c$ from \gls{PPC}, calibrated \gls{PPC}s \citep{robins2000asymptotic,Hjort:2006}, \gls{HPC} and prior PC. (b) mean squared error of fitted model (\Cref{eq:ridge_model}) over different values of $c$.} \label{fig:regression_main}
\end{figure}

As anticipated, the \gls{PPC} $p$-values are constant for all values of the regularization parameter, $c$; that is, the \gls{PPC} cannot detect model misfit for large values of $c$ (\Cref{fig:regression_main}(b)). Moreover, the two calibrated \gls{PPC}s \citep{robins2000asymptotic,Hjort:2006} also cannot detect this model misfit. Meanwhile, the \gls{HPC} retains the model for small values of $c$. For large values of $c$, the \gls{HPC} rejects the model as it overfits to the observed data. The prior predictive check exhibits similar behavior to the \gls{HPC}, but it does not begin to reject the model until larger values of $c$.

\begin{figure}
\begin{subfigure}[t]{0.5\textwidth}
	\centering
		\includegraphics[width=\textwidth]{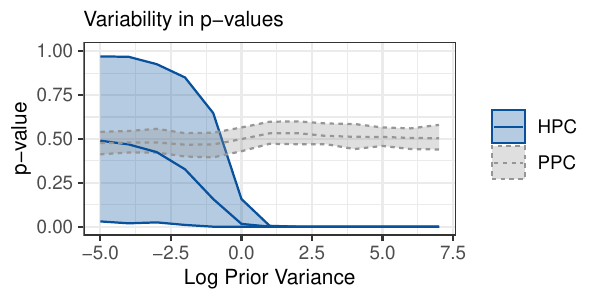}
\caption{\gls{HPC} vs. \gls{PPC}}\label{fig:pop_pc_variability}
\end{subfigure}
\begin{subfigure}[t]{0.5\textwidth}
	\centering
		\includegraphics[width=\textwidth]{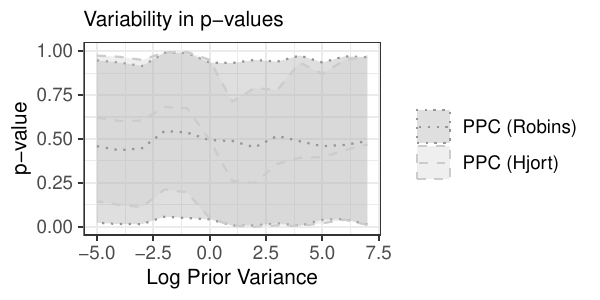}
\caption{Calibrated \glspl{PPC} \citep{robins2000asymptotic,Hjort:2006}}\label{fig:ppc_variability}
\end{subfigure}
	\caption{Bayesian ridge regression: (a) When the model is not overfit, the \gls{HPC} $p$-values are much more variable than \gls{PPC} $p$-values, which concentrate around 0.5. When the model is overfit, \gls{HPC} $p$-values detect this overfitting while \gls{PPC} $p$-values do not. (b) Calibrated \gls{PPC} $p$-values are much more variable than the \gls{PPC}; however, the calibrated \gls{PPC} does not detect overfitting for large $c$.  For both (a) and (b), lower and upper lines show 0.025 and 0.975 quantiles of $p$-values over $K=100$ replicates for different values of the regularization parameter $c$; middle line shows mean.} 
\end{figure}

\begin{figure}[h]
\begin{subfigure}[t]{0.49\textwidth}
\centering
\includegraphics[width = \textwidth]{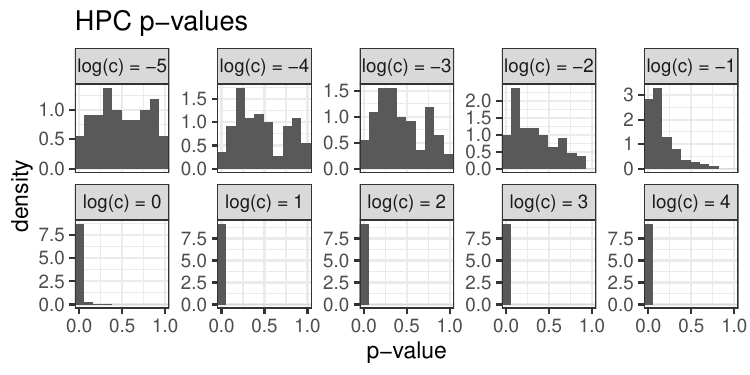}
\end{subfigure}
\begin{subfigure}[t]{0.49\textwidth}
\centering
\includegraphics[width = \textwidth]{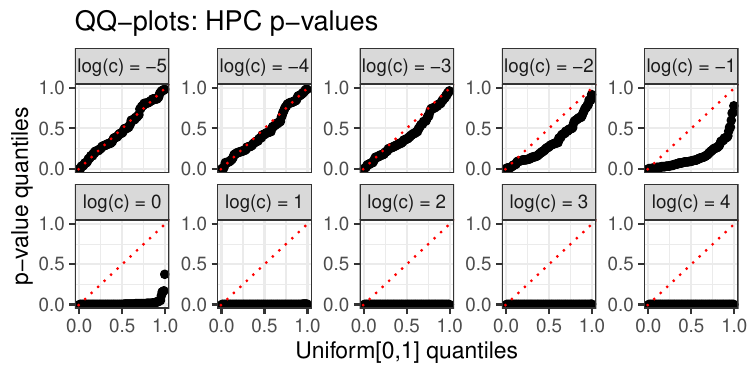}
\end{subfigure}
\caption{Bayesian ridge regression: \gls{HPC} $p$-values are approximately uniform when there is no overfitting (small $c$); then, the \gls{HPC} detects when there is overfitting (large $c$). Left: Histograms of \gls{HPC} $p$-values. Right: QQ-plots comparing the quantiles of the \gls{HPC} $p$-values with the quantiles of a uniform[0,1] random variable. Plots are over $K=100$ replications of the data.} \label{fig:pop_pc_p_vals}
\end{figure}

We next consider the variability of the $p$-values from the predictive checks. As anticipated by the theoretical results of \citet{robins2000asymptotic}, the \gls{PPC} $p$-value is tightly concentrated around 0.5 for all values of $c$ (\Cref{fig:ppc_variability}).  Meanwhile, the empirical calibration procedures of \citet{robins2000asymptotic,Hjort:2006} give $p$-values with greater variability around 0.5, as expected from the calibration process; however, they still do not reject the model for large $c$ (\Cref{fig:ppc_variability}). In contrast, the \gls{HPC} $p$-values show greater variability around 0.5 for small values of $c$ and then concentrate around 0 for large values of $c$ (\Cref{fig:pop_pc_variability}).  

The empirical distribution of the \gls{HPC} $p$-values are displayed in \Cref{fig:pop_pc_p_vals}. The  $p$-values are approximately uniform for small values of $c$. This provides empirical evidence for the calibration of \gls{HPC} $p$-values with the $\chi^2$-diagnostic. For large values of $c$, the $p$-values concentrate around 0; that is, the \gls{HPC} has high power to detect model misfit.  Additional histograms and QQ-plots of the $p$-value distributions for all methods are in \Cref{appendix:experiments}.

In theory, the \gls{HPC} is calibrated for asymptotically normal diagnostic statistics (\Cref{theorem:pop_pc_ideal_calibration}). In this regression example, the diagnostic statistic is $\chi^2$ distributed, not normal. Although our theory does not extend to this diagnostic, we showed empirically that the \gls{HPC} has good calibration properties.

\subsection{Topic Modeling}

We next study \acrshortpl{HPC} on \gls{LDA} \citep{Blei:2003b}, a
hierarchical model of documents. We first apply LDA to synthetic data and show empirically the \gls{HPC} has good calibration properties. We then apply LDA to a collection of documents from the \emph{New York Times} and show that the \gls{HPC} detects model misfit due to overfitting.

\acrshort{LDA} models documents as mixtures over
latent topics, where each topic is a distribution over words. The $k$th topic is denoted by 
$\beta_k=\{\beta_{kv}\}_{v=1}^V$, where $V\in\mathbb{N}$ is the number of unique words in the corpus, and the number of topics ranges from $k=1,\dots, K$. The topics are drawn as:
\begin{align}
\beta_k \sim \text{Dirichlet}(\alpha \mathbf{1}_V), \quad k = 1,\dots, K,
\end{align}
where $\alpha\in\mathbb{R}^+$ is a hyperparameter.

For the $d$th document, the generative process is:
\begin{enumerate}
	\item Draw the topic proportions $\theta_d \sim \text{Dirichlet}(\alpha \mathbf{1}_K) $
	\item For words $n=1,\dots, N_d$:
	\begin{enumerate}
		\item Draw a topic $z_{dn} \sim \text{Multinomial}(\theta_d)$
		\item Draw a word conditioned on the topic $w_{dn} \sim \text{Multinomial}(\beta_{z_{dn}})$
	\end{enumerate}
\end{enumerate}

We set the Dirichlet hyperparameter to $\alpha=0.1$ on both the topics and the document proportions.  

For the model check diagnostic, we will use the log-likelihood.  For document $d$, the log-likelihood is:
\begin{align}
\ell (\bm{w}_d; \theta_d, \bm{\beta}) = \sum_{v=1}^V \left[\sum_{n=1}^{N_d} \mathbbm{1}(w_{dn}=v)\right] \log\left(\sum_{k=1}^K \beta_{kv}\theta_{dk}\right).
\end{align}

To calculate the \gls{PPC} $p$-value, we first calculate the posterior expectation of the global topics $\widehat{\bm{\beta}} = \mathbb{E}[\bm{\beta}|\widetilde{\mbw}]$ where $\widetilde{\mbw}$ is a heldout set of documents. Then, we estimate the local per-document parameters, $\{\theta_d\}_{d=1}^D$, and draw replicates from the posterior predictive distribution. The per-document \gls{PPC} is:
\begin{align}
	\textrm{ppc}(\mbw_d^{\textrm{obs}}) &= \frac{1}{R}\sum_{r=1}^{R} \mathbbm{1}[ \ell(\mbw_{d, r}^{\textrm{rep}}, \theta_{d,r}, \widehat{\bm{\beta}}) > \ell(\mbw_d^{\textrm{obs}}, \theta_{d,r}, \widehat{\bm{\beta}})], \\
\text{where} &\quad (\mbw_{d, r}^{\textrm{rep}}, \theta_{d,r}) \sim p(\mbw_d^{\textrm{rep}}|\theta_d, \widehat{\bm{\beta}})p(\theta_d|\mbw_d^{\textrm{obs}}, \widehat{\bm{\beta}}).
\end{align}
Note that the topic vector $\bm{\beta}$ is fixed as the estimate from the heldout corpus; the above PPC is checking the fit of the model based on the document topic proportions $\theta_d$.

To calculate the \gls{HPC} $p$-value, we split each document in half: $\mbw_d = \{\mbw_d^{\textrm{obs}}, \mbw_d^{\textrm{new}}\}_{d=1}^D$. Here, we split the data at the document level because for each document, we need to infer the local variable, $\theta_d$. Half of the document is used to infer this per-document variable, while the other half is used to check the model:
\begin{align}
	\textrm{hpc}(\mbw_d^{\textrm{obs}}, \mbw_d^{\textrm{new}}) &= \frac{1}{R}\sum_{r=1}^{R} \mathbbm{1}[ \ell(\mbw_{d, r}^{\textrm{rep}}, \theta_{d,r}, \widehat{\bm{\beta}}) > \ell(\mbw_d^{\textrm{new}}, \theta_{d,r}, \widehat{\bm{\beta}})], \\
\text{where} &\quad (\mbw_{d, r}^{\textrm{rep}}, \theta_{d,r}) \sim p(\mbw_d^{\textrm{rep}}|\theta_d, \widehat{\bm{\beta}})p(\theta_d|\mbw_d^{\textrm{obs}}, \widehat{\bm{\beta}}).
\end{align}

\subsection{Synthetic data}

We investigate the distribution of \gls{HPC} and \gls{PPC} $p$-values when the data is drawn from the LDA generative process with $K=10$ topics and $V=2500$ vocabulary of unique words. The number of words (tokens) in each document is drawn as $N_d\sim \text{Poisson}(\xi)$ where $\xi=300$.  

We draw $100,000$ documents and infer the topics $\beta$ using stochastic variational inference \citep{Hoffman:2013} with minibatches of size
100.   Given the topics $\beta$, we then compare the \gls{PPC} and \gls{HPC} on a heldout set of documents of size $D=1,000$.

\begin{figure}
\centering
	\includegraphics[width=0.8\textwidth]{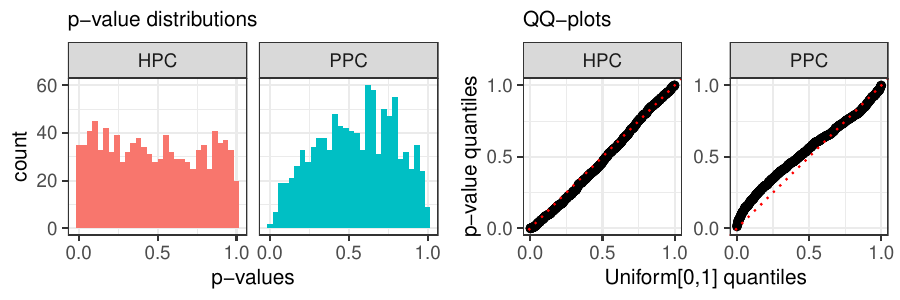}
	\caption{Simulated LDA data: when the model is true, \gls{HPC} $p$-values are approximately uniform while \gls{PPC} $p$-values are not. Left: Histograms of \gls{HPC} and \gls{PPC} $p$-values. Right: QQ-plots comparing the quantiles of the \gls{HPC} and \gls{PPC} $p$-values with the quantiles of a uniform[0,1] random variable.} \label{fig:topic_sim}
\end{figure}

For each document, we calculate the \gls{PPC} and \gls{HPC} $p$-values using the log-likelihood diagnostic with $R=500$ replicates from the posterior predictive distribution. The \gls{HPC} $p$-values are approximately uniformly distributed while the \gls{PPC} $p$-values are left-skewed (\Cref{fig:topic_sim}). This provides empirical evidence that the \gls{HPC} is calibrated for the LDA log-likelihood diagnostic.

\subsection{New York Times}

We consider a corpus of 100,000 news documents with a vocabulary size of 5,000 unique words from \emph{the New York Times}.  To infer the topics $\beta$, we implement stochastic variational inference~\citep{Hoffman:2013} with minibatches of size
100.   Given the topics $\beta$, we then compare the \gls{PPC} and \gls{HPC} on a set of documents of size 1,000.  

For each document, we calculate the \gls{PPC} and \gls{HPC} $p$-values using the log-likelihood diagnostic averaged over documents, where for each document we draw $R=500$ replicates from the posterior predictive distribution.  

\begin{figure}
\centering
	\includegraphics[width=0.75\textwidth]{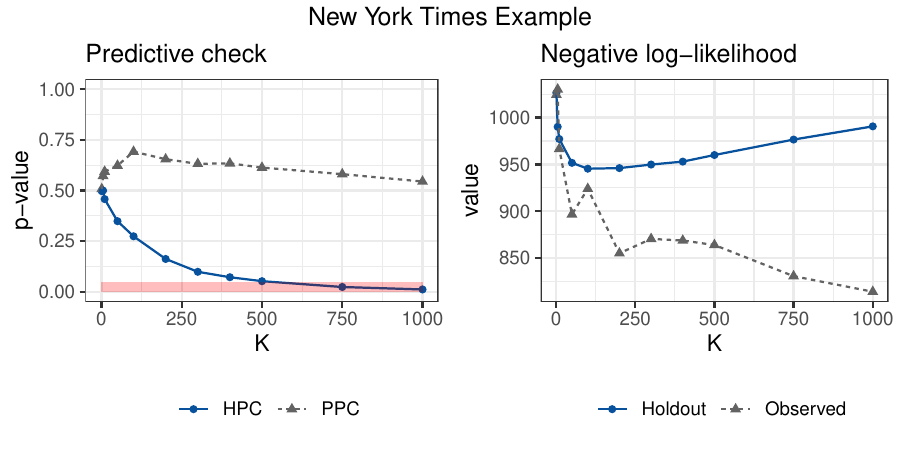}
	\caption{On the New York Times data, the \gls{HPC} detects model overfitting, rejecting the model when the number of topics is large. In contrast, the \gls{PPC} never rejects the model. Left: \gls{HPC} and \gls{PPC} $p$-values for different values of topics $K$. Right: Negative log-likelihood on observed and holdout data over different values of topics $K$.} \label{fig:nyt}
\end{figure}

As the number of topics increases to $K=1,000$, the \gls{PPC} $p$-value remains close to 0.5, and the negative log-likelihood is monotonically decreasing (\Cref{fig:nyt}). At $K=1,000$, the number of topics is equal to the size of the vocabulary; at this stage, the model assigns each word to its own topic, memorizing the data. The \gls{HPC} $p$-value detects this overfitting, and begins to reject the model past $K=200$ topics (\Cref{fig:nyt} left). 

Again, we are not using the \gls{HPC} to choose the number of topics $K$; we are simply emphasizing that when there is clear overfitting, the \gls{HPC} is able to detect this poor model fit, while the \gls{PPC} cannot.

Note that when the number of topics is small, the \gls{HPC} does not reject the model. A similar phenomenon was noted in \citet{moran2021posterior} for a Gaussian mixture model when the number of mixtures components is fewer than the truth. When the number of mixture components is too few, the entropy of the posterior predictive distribution increases to fit the data. The draws from this posterior distribution can be extreme relative to the true model, making it difficult to distinguish model misfit.

\subsection{Factor Analysis}

In this section, we study the \gls{HPC} for factor analysis, using examples similar to \citet{moran2021posterior}.  We fit probabilistic principal component analysis \citep[PPCA,][]{TB99} to (i) data drawn from a well-specified linear model and (ii) data drawn from a nonlinear model for which PPCA is not well-specified.  We show empirically that:
\begin{itemize}
\item \gls{PPC} $p$-values are not calibrated and have low power to detect the poor fit of the nonlinear model;
\item \gls{HPC} $p$-values, while not exactly calibrated, avoid the degeneracy of the \gls{PPC} $p$-values when the model is correct. When the model is incorrect, the \gls{HPC} has much higher power than the \gls{PPC}.
\end{itemize}

\parhead{Data generating process}

The observed data is $\mbx_i \in \RR^G$, $i = 1,\dots, N$. We assume that $\bm{x}_i$ has some low dimensional representation $\bm{z}_i \in \RR^K$ with 
\begin{align}
	\bm{x}_i = f(\bm{z}_i) + \bm{\varepsilon}_i \label{eq:factor_analysis_model}
\end{align}
for some function $f: \RR^K \to \RR^G$ and noise term $\bm{\varepsilon}_i \in \RR^G$.  We will consider two cases: (i) $f$ is set to a linear function and (ii) $f$ is set to a nonlinear function.

\parhead{Model}

Probabilistic PCA \citep{TB99} assumes that $f$ is a linear mapping from the low-dimensional latent representation to the observed data,
\begin{align*}
	\bm{x}_i = \bm{W} \bm{z}_i + \bm{\varepsilon}_i, \quad \bm{\varepsilon}_i \sim N(\bm{0}, \sigma^2\bm{I}).
\end{align*}
The latent variables are assigned a normal prior, $\bm{z}_i \sim N(0, \bm{I})$.  We fit $\bm{W}$ and representations $\bm{z}_i$ using the EM algorithm.

According to \Cref{eq:factor_analysis_model}, when $f$ is linear, PPCA should adequately fit the data. In contrast, when $f$ is nonlinear, PPCA does not have the requisite complexity to fit the data.

\parhead{Diagnostic function}

To check the model, we use the likelihood diagnostic. For PPCA, this is:
\begin{align}
	d_{\mathrm{ppca}}(\mbx;\mbx_{\mathrm{obs}}) &= \sum_{i=1}^N  \frac{1}{2\widehat{\sigma^2}}\lVert\mbx_i - \widehat{\mbW}\mathbb{E}[\mbz|\mbx_i, \widehat{\mbW}]\rVert^2, \quad \text{where} \quad (\widehat{\mbW}, \widehat{\sigma}^2)=\argmax_{\mbW, \sigma^2}\log p(\mbx_{\obs}|\mbW, \sigma^2)\\
&\text{and} \quad \mathbb{E}[\mbz_i|\mbx_i, \widehat{\mbW}, \widehat{\sigma}^2]=\mbM^{-1}\widehat{\mbW}^T\mbx_i, \quad \mbM=\widehat{\mbW}^T\widehat{\mbW}+ \widehat{\sigma}^2\mbI.
\end{align}

Then, the empirical \gls{PPC} and \gls{HPC} are:
\begin{align}
p_{\ppc} &= \frac{1}{R}\sum_{r=1}^R \mathbb{I}[d(\mbx_{\rep};\mbx_{\mathrm{obs}}) > d(\mbx_{\obs};\mbx_{\mathrm{obs}})];\\
p_{\hpc} &= \frac{1}{R}\sum_{r=1}^R \mathbb{I}[d(\mbx_{\rep};\mbx_{\mathrm{obs}}) > d(\mbx_{\new};\mbx_{\mathrm{obs}})].
\end{align}

\subsubsection{Linear data generating process} \label{sec:linear-example}

We first consider the setting where $f$ is linear.  We set the number of samples to  $N=1000$, the number of observed features to $G=11$ and the latent dimension as $K=6$. The data is generated as
\begin{align*}
	\bm{x}_i = \bm{W}\bm{z}_i + \bm{\varepsilon}_i, 
\end{align*}
where $\bm{z}_i \sim N(0, \bm{I})$, $\bm{\varepsilon}_i\sim N(0, \sigma^2\bm{I})$ with true $\sigma^2=1$.  (Note however that $\sigma^2$ is treated as unknown in the inference stage). The matrix $\bm{W}$ is the matrix,
\begin{align*}
\setcounter{MaxMatrixCols}{11}
	\bm{W}^{\top} = 
	\begin{pmatrix}
		5 & 5 & 5 & 0 & 0 & 0 & 0 & 0 & 0 & 0 & 0 \\
		0 & 0 & 5 & 5 & 5 & 0& 0 & 0 & 0 & 0 & 0 \\
		0 & 0 & 0 & 0 & 5 & 5& 5 & 0 & 0 & 0 & 0 \\
		0 & 0 & 0 & 0 & 0 & 0& 5 & 5 & 5 & 0 & 0 \\
		0 & 0 & 0 & 0 & 0 & 0& 0 & 0 & 5 & 5 & 5 \\
	\end{pmatrix}.
\end{align*}

\begin{figure}[!t]
\centering
	\includegraphics[width=0.8\textwidth]{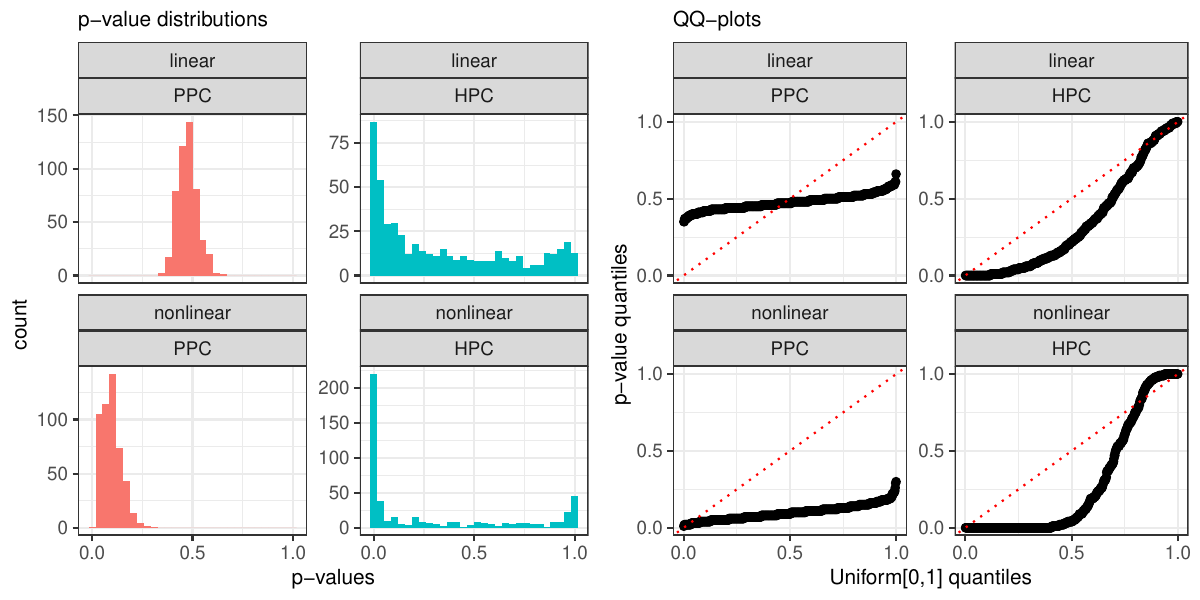}
	\caption{Factor analysis example. Top row: when the model is correctly specified (linear), the \gls{HPC} avoids the degeneracy of the \gls{PPC} around 0.5. Bottom row: when the model is incorrect (nonlinear), the \gls{HPC} has higher power than the \gls{PPC} to reject the model.} \label{fig:factor_sim}
\end{figure}

In this setting, the \gls{PPC} is centered around 0.5 (\Cref{fig:factor_sim}, top row). Meanwhile, the \gls{HPC}, while not perfectly calibrated, avoids this degeneracy issue of the \gls{PPC}.

\subsubsection{Nonlinear data generating process} \label{sec:nonlinear-example}

We next consider the setting where the true mapping $f$ from the factors to the observed data is nonlinear. Here, we expect PPCA to be a poor fit for the nonlinear model. We set the number of samples to  $N=1000$, the number of observed features to $G=6$ and the latent dimension to $K=2$. The data is generated from
\begin{align}
	\mbx_i = (z_{i1},\ 2z_{i1},\ 3z_{i1}^2, \ 4z_{i2},\ 5z_{i2},\ 6\sin(\pi/2 \cdot z_{i2}))^{\top} + \boldsymbol{\epsilon}_i, \label{eq:nonlinear}
\end{align}
where $\bm{z}_i \sim N(0, \bm{I})$, $\bm{\varepsilon}_i\sim N(0, \sigma^2\bm{I})$ with true $\sigma^2=1$.  That is, the first three columns of $\mbx$ are related to the first factor and the next three columns are related to the second factor.

In this setting, the \gls{PPC} $p$-values have a median of 0.09 and reject the model only 12.6\% of the time (\Cref{fig:factor_sim}, bottom row). Meanwhile, the \gls{HPC} rejects the model 63.2\% of the time (\Cref{fig:factor_sim}, bottom row). That is, the \gls{HPC} empirically has much higher power than the \gls{PPC} to detect the inadequacy of PPCA for modeling the nonlinear data generating process.


%% file: sec_discussion.tex

\glsresetall

\section{Discussion}
\label{sec:discussion}
\glsresetall

We developed \glspl{HPC}, a diagnostic tool that brings together
Bayesian methods for model checking with frequentist estimation of
goodness of fit. The \gls{HPC} assesses a Bayesian model by comparing
samples from the posterior predictive to a sample from the population
distribution, which in practice is a holdout dataset. We proved that
the \gls{HPC} is calibrated for a class of asymptotically normal
diagnostic functions and empirically show its calibration for other
diagnostics. We revisited the ``double use of the data'' issue of
\glspl{PPC} and highlighted that while post-hoc procedures can be used
to calibrate the \gls{PPC}, post-hoc calibration does not provide
power to detect model misfit.  Finally, we demonstrated the utility of
\glspl{HPC} with Bayesian linear regression models, on
probabilistic topic models of documents, and on factor analysis.

There are several areas for further research.
For hierarchical models of grouped data, \citet{Marshall:2003} define
mixed predictive checks, where the reference distribution combines the
prior for the group-specific latent variables with the posterior for
latent variables shared across groups. This approach mitigates the
 issues of a \gls{PPC}, but there are no guarantees; a mixed
predictive check can still be uncalibrated or have low power if the influence of the
posterior becomes too large. To avoid these issues,
\citet{Bayarri:2007} extends the checks of \citet{bayarri2000p} to
hierarchical models.  How to extend the \gls{HPC} to these situations is
one avenue of further research.

Generalizing beyond hierarchical models, researchers have studied how
to check individual components of a probabilistic model, i.e.,
individual nodes in a directed graphical model.  \citet{OHagan:2003}
proposes some of the earlier ideas along these lines, though in a way
that uses the data twice and leads to an uncalibrated check. To correct
this lack of calibration, his predictive checks for individual
components of a model have been extended, often by using data
splitting~\citep{Marshall:2007,Bayarri:2007,Dahl:2007,Gasemyr:2009,Presanis:2013}.
Again the \gls{HPC} proposed here could be extended to these settings.


%% file: appendix.tex
\section{Proofs}

\subsection{Proof of calibration of the holdout predictive check} \label{appendix:calibration_proof}

In this section, we prove \Cref{theorem:pop_pc_ideal_calibration}.  Recall that we assume the diagnostic $d(\mby)$ is asymptotically normal with asymptotic mean $\nu(\theta)$ and asymptotic variance $\sigma^2(\theta)$, under the null hypothesis that the density of $\mby$ is $f(\mby;\theta)$:
\begin{align}
n^{1/2}\left[\frac{d(\mby) - \nu(\theta)}{\sigma(\theta)}\right]\leadsto N(0,1),\label{eq:pop_pc_a1}
\end{align}
where $\leadsto$ denotes convergence in distribution.

{\bf Notation.}
We let $\rmp(\theta|\mby)$ denote the posterior distribution of $\theta$. We use $\lVert \cdot \rVert$ to denote the total variation distance between two distributions $P$ and $Q$. The Fisher information is
\begin{align}
I(\theta) = \lim_{n\to\infty} n^{-1} \mathbb{E}_{\theta}\left[-\partial^2 \log f(\mby; \theta)/\partial\theta\partial\theta'\right].
\end{align}
We use the notation $y_n = O_p(a_n)$ as $n\to\infty$ to mean $y_n/a_n$ is stochastically bounded: for any $\varepsilon > 0$, there exists finite $M>0$ and $N>0$ such that
\begin{align}
\rmp(|y_n/a_n| > M) < \varepsilon, \quad \forall n > N.
\end{align}
We use $\phi(\theta;\mu, \Sigma)$ to denote the density of a $N(\mu, \Sigma)$ random variable.

{\bf Regularity conditions.}
\begin{enumerate}
\item The asymptotic mean $\nu$ is continuously differentiable in a neighborhood of $(0,\theta)$, with partial derivatives converging to limit:
\begin{align}
\dot{\nu}(\theta_0) &= \lim_{n\to\infty} \partial \nu_n(0, \theta)/\partial\theta \big |_{\theta=\theta_0}.
\end{align}
\item For some $p$-vector-valued function $\theta(\mby)$ on the sample space, we assume
\begin{align}
\lVert \rmp(\cdot | \mby) - N(\theta(\mby), I^{-1}(\theta_0)/n) \rVert \stackrel{P_{\theta_0}}{\to} 0 \label{eq:calibration_condition_2}
\end{align}
and
\begin{align}
n^{1/2}(\theta(\mby) - \theta_0) = O_{P_{\theta_0}}(1). \label{eq:calibration_condition_3}
\end{align}
\end{enumerate}

\paragraph{Proof of \Cref{theorem:pop_pc_ideal_calibration}.}
Our proof technique follows that of \citet{robins2000asymptotic} for their proof of Theorem 3.

The \gls{HPC} $p$-value is:
\begin{align}
p(\yobs, \ynew) = \int_{\Theta} \rmp(d(\yrep) > d(\ynew) | \ynew, \theta)\rmp(\theta|\yobs) d\theta \label{eq:pop_pc_start_theorem_proof}
\end{align}

Consider:
\begin{align}
d(\yrep) &> d(\ynew) \\ 
\text{Then,}\qquad\qquad \sqrt{n}[d(\yrep) - \nu(\theta)] &> \sqrt{n}[d(\ynew) - \nu(\theta_0) - (\nu(\theta) - \nu(\theta_0))]. \label{eq:pop_pc_centered_1} 
\end{align}

Consider the LHS of  \Cref{eq:pop_pc_centered_1}. By \Cref{eq:pop_pc_a1}, we have that, conditional on $\theta$,
 $$\sqrt{n}[d(\yrep) - \nu(\theta)]  \sim N(0, \sigma^2(\theta_0)).$$

Consider now the RHS of \Cref{eq:pop_pc_centered_1}.  The first term $\sqrt{n}[d(\ynew) - \nu(\theta_0)]$ is fixed, conditional on $\ynew$. For the second term, we use a Taylor expansion to obtain:
\begin{align}
\nu(\theta) - \nu(\theta_0) \approx \dot{\nu}(\theta_0)(\theta - \theta_0).
\end{align}
Then by \Cref{eq:calibration_condition_2}, given $\yobs$, the second term is approximately distributed as 
\begin{align}
N(\dot{\nu}(\theta_0)^T n^{1/2}(\theta(\yobs) - \theta_0), \dot{\nu}(\theta_0)^T I^{-1}(\theta_0) \dot{\nu}(\theta_0)).
\end{align}
(Note this step using \Cref{eq:calibration_condition_2} helps resolve the potentially difficult integral over $\rmp(\theta|\yobs)$.)

Then, the conditional probability, given $\yobs$ and $\ynew$, of $d(\yrep) > d(\ynew)$ is approximately
\begin{align}
\rmp(W > 0 | \yobs, \ynew), 
\end{align}
where $W$ is a Gaussian random variable with  
\begin{align}
\mathbb{E}[W|\yobs, \ynew] &= n^{1/2}\dot{\nu}(\theta_0)^T  (\theta(\yobs) - \theta_0)-n^{1/2}[d(\ynew) - \nu(\theta_0)],\\
\text{Var}(W|\yobs, \ynew) &= \sigma^2(\theta_0) + \dot{\nu}(\theta_0)^T I^{-1}(\theta_0) \dot{\nu}(\theta_0).
\end{align}

Then, the \gls{HPC} $p$-value is
\begin{align}
p(\yobs, \ynew)  &= 1 - \Phi\left( \frac{n^{1/2} \dot{\nu}(\theta_0)^T  (\theta(\yobs) - \theta_0)-n^{1/2}[d(\ynew) - \nu(\theta_0)]}{\sqrt{ \sigma^2(\theta_0) + \dot{\nu}(\theta)^T I^{-1}(\theta_0) \dot{\nu}(\theta)}}\right) + o_{P_0}(1),
\end{align}
where we have used \Cref{eq:calibration_condition_2}.

Now, we consider the distribution of the \gls{HPC} $p$-value over the distributions of $\yobs$ and $\ynew$.

By \Cref{eq:calibration_condition_3}, we have that the posterior mean converges to the true $\theta_0$:
\begin{align}
\sqrt{n}\dot{\nu}(\theta_0)^T (\theta(\yobs) - \theta_0) \leadsto N(0, \dot{\nu}(\theta_0)^T I^{-1}(\theta_0) \dot{\nu}(\theta_0)).
\end{align} 
Independently, $n^{1/2}[d(\ynew) - \nu(\theta_0)]$ converges to a $N(0, \sigma^2(\theta_0))$ distribution.

 Hence, the  \gls{HPC} $p$-value is:
\begin{align}
p(\yobs, \ynew)&= 1 - \Phi\left( Q\right) + o_{P_0}(1),
\end{align}
where $Q\sim N(0, 1)$.

This concludes the proof.

\section{Note on previous version of this paper} \label{appendix:pop_pc_v1}

The previous version of this paper, \citet{ranganath2019population}, proposed the \gls{PoPC}. The \gls{PoPC} treated $\ynew$ as random; in the current holdout predictive check, $\ynew$ is fixed. Treating $\ynew$ as fixed results in a calibrated check, while treating $\ynew$ as random does not. For completeness, we include the previous definition of the \gls{PoPC} below.

\begin{definition}[Population predictive check (Version 1, 2019)] 
Consider observed data $\yobs$, its posterior predictive distribution $\rmp(\yrep \g \yobs)$, and a diagnostic statistic $d(\mby)$. Suppose we have $\ynew$ drawn from the population distribution of the data. The population predictive check as a $p$-value is:
\begin{align}
  p_{\mathrm{pop-v1}} = \rmp(d(\yrep) \geq d(\ynew) \g \yobs), \quad \ynew \sim F\label{eq:old-pop-pc-def}
\end{align}
where $\yrep \sim \rmp(\yrep \g \yobs)$.
\end{definition}

To illustrate the issue with treating $\ynew$ as random, we again consider the mean example in \Cref{sec:problem_with_ppcs}. Suppose we observe data $\yobs = \{y_i\}_{i=1}^n$ drawn from a Gaussian with known variance parameter:
\begin{align}
y_i  \sim N(\mu, \sigma^2),\label{eq:h0_mean_old}
\end{align}
for some $\mu\in\mathbb{R}$ and fixed $\sigma^2$. For a prior on $\mu$, we take $\mu \sim N(\mu_0, \sigma_0^2)$ with fixed hyperparameters $\mu_0\in\mathbb{R}, \sigma_0^2\in\mathbb{R}^+$. 

\begin{figure}
\centering
\includegraphics[width =0.3\textwidth]{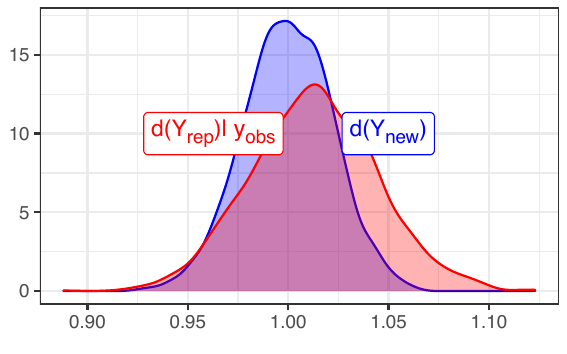}
\caption{The distribution of $d(\ynew)=\overline{\mby}^\new$ is not equal to the distribution of $d(\yrep)=\overline{\mby}^\rep$. (Data drawn as $y_i\sim N(1, 1)$, $i=1,\dots, 2000$). }\label{fig:mean_old_pop_pc}
\end{figure}

In this case, the posterior predictive distribution is 
\begin{align}
\overline{\mby}^\rep | \overline{\mby}^\obs \sim N\left(\rho_n \overline{\mby}^\obs + (1 - \rho_n)\mu_0, \left(1 + \rho_n\right)\frac{\sigma^2}{n} \right).
\end{align}
As $n\to\infty$, $\overline{\mby}^\rep | \overline{\mby}^\obs$ is centered around $\overline{\mby}^\obs$. This is different from the distribution of $\ynew$, which is the population distribution: $\overline{\mby}^\new\sim N(\mu, \sigma^2)$ (see \Cref{fig:mean_old_pop_pc}). Consequently, the \acrshort{PoPC} is not calibrated. We demonstrate this lack of calibration theoretically below.

Consider the distribution of \acrshort{PoPC} when \Cref{eq:h0_mean_old} holds:
\begin{align}
\rmp(d(\yrep) > d(\ynew)| \yobs) = \rmp(D > 0 | \yobs),
\end{align}{}
where $D = d(\yrep) - d(\ynew)$ is a Gaussian random variable with 
\begin{align}
\E[D|\yobs] = \rho_n\overline{\mby}^\obs+ (1 - \rho_n)\mu_0 - {\mu}, \quad Var(D) = \left(2 + \frac{\rho_n}{n}\right)\frac{\sigma^2}{n}.
\end{align}
The \acrshort{PoPC} is then:
\begin{align} 
\rmp(D > 0 | \yobs) = 1 - \Phi\left( \frac{{\mu} - \rho_n \overline{\mby}^\obs - (1 - \rho_n)\mu_0}{\sqrt{(2 + \rho_n) \sigma^2/n}}\right)
\end{align}
Now, if the observed data has the same distribution as the new data, we have $\overline{\mby}^\obs\sim N({\mu}, \sigma^2/n)$. Then, for large $n$, the \gls{PoPC} $p$-value (Version 1, 2019) is:
\begin{align}
\rmp(D > 0 | \yobs)  \to 1 -  \Phi\left( Z/\sqrt{3} \right), \quad \text{where } Z \sim N(0, 1).
 \end{align}
Consequently, the \acrshort{PoPC} is not calibrated.

The updated definition of the \gls{HPC} in \Cref{def:pop-pc} does not have this calibration issue. Intuitively, this is because a fixed $\ynew$ is on average as far from the true mean as the fixed $\yobs$. 

For the more general case of asymptotically normal diagnostic functions, we also prove the \acrshort{PoPC} is not uniformly distributed.

\begin{theorem}\label{theorem:old_pop_pc_distribution}
We assume \Cref{eq:pop_pc_diagnostic_convergence} holds, in addition to regularity conditions detailed in \Cref{appendix:calibration_proof}. Under the distribution $f(\mby; \theta_0)$, the \acrshort{PoPC} $p$-value can be written as:
\begin{align}
p_{\mathrm{pop}}(\mby) = 1 - \Phi(Q)+o_P(1), \text{ where } Q \sim N\left(0, \frac{\dot{\nu}_{\theta}(\theta_0)^T I^{-1}(\theta_0) \dot{\nu}_{\theta}(\theta_0)}{2\sigma^2(\theta_0) + \dot{\nu}_{\theta}(\theta_0)^T I^{-1}(\theta_0) \dot{\nu}_{\theta}(\theta_0)} \right),
\end{align}
where $o_P(1)$ denotes a random variable converging to zero in probability, $\Phi$ is the standard normal cdf, and $Q\sim N(0,1)$.

Consequently, the \acrshort{PoPC} is not calibrated.
\end{theorem}

\paragraph{Proof of \Cref{theorem:old_pop_pc_distribution}.}

The \acrshort{PoPC} $p$-value is:
\begin{align}
p(\yobs) = \int_{\Theta} \rmp(d(\yrep) > d(\ynew) |  \theta)\rmp(\theta|\yobs) \rmp(\ynew|\theta_0) d\theta \label{eq:pop_pc_v1_start_theorem_proof}
\end{align}

Consider:
\begin{align}
d(\yrep) &> d(\ynew) \\ 
\text{Then,}\qquad\qquad \sqrt{n}[d(\yrep) - \nu(\theta)] &> \sqrt{n}[d(\ynew) - \nu(\theta_0) - (\nu(\theta) - \nu(\theta_0))]. \label{eq:pop_pc_v1_centered_1} 
\end{align}

Consider the LHS of  \Cref{eq:pop_pc_v1_centered_1}. By \Cref{eq:pop_pc_a1}, we have
 $$\sqrt{n}[d(\yrep) - \nu(\theta)]  \sim N(0, \sigma^2(\theta_0)).$$

Consider now the RHS of \Cref{eq:pop_pc_v1_centered_1}.  The first term is distributed as:
\begin{align}
\sqrt{n}[d(\ynew) - \nu(\theta_0)] \sim N(0, \sigma^2(\theta_0)).
\end{align}
For the second term, we use a Taylor expansion to obtain:
\begin{align}
\nu(\theta) - \nu(\theta_0) \approx\dot{\nu}(\theta_0)(\theta - \theta_0).
\end{align}
Then by \Cref{eq:calibration_condition_2}, given $\yobs$, the second term is approximately distributed as 
\begin{align}
N(\dot{\nu}(\theta_0)^T n^{1/2}(\theta(\yobs) - \theta_0), \dot{\nu}(\theta_0)^T I^{-1}(\theta_0) \dot{\nu}(\theta_0)).
\end{align}
(Note this step using \Cref{eq:calibration_condition_2} helps resolve the potentially difficult integral over $\rmp(\theta|\yobs)$.)

Then, the conditional probability, given $\yobs$, of $d(\yrep) > d(\ynew)$ is approximately
\begin{align}
\rmp(W > 0 | \yobs), 
\end{align}
where $W$ is a Gaussian random variable with  
\begin{align}
\mathbb{E}[W|\yobs] &= n^{1/2}\dot{\nu}(\theta_0)^T (\theta(\yobs) - \theta_0),\\
\text{Var}(W|\yobs) &= 2\sigma^2(\theta_0) + \dot{\nu}(\theta_0)^T I^{-1}(\theta_0) \dot{\nu}(\theta_0).
\end{align}

Then, the \gls{PoPC} $p$-value is
\begin{align}
p(\yobs)  &= 1 - \Phi\left( \frac{n^{1/2} \dot{\nu}(\theta_0)^T  (\theta(\yobs) - \theta_0)}{\sqrt{ 2\sigma^2(\theta_0) + \dot{\nu}(\theta)^T I^{-1}(\theta_0) \dot{\nu}(\theta)}}\right) + o_{P_0}(1),
\end{align}
where we have used \Cref{eq:calibration_condition_2}.

Now, we consider the distribution of the \acrshort{PoPC} $p$-value over the distribution of $\yobs$.

By \Cref{eq:calibration_condition_3}, we have that the posterior mean converges to the true $\theta_0$:
\begin{align}
\sqrt{n}\dot{\nu}(\theta_0)^T (\theta(\yobs) - \theta_0) \leadsto N(0, \dot{\nu}(\theta_0)^T I^{-1}(\theta_0) \dot{\nu}(\theta_0)).
\end{align} 

 Hence, the  \acrshort{PoPC} $p$-value is:
\begin{align}
p(\yobs)&= 1 - \Phi\left( Q\right) + o_{P_0}(1),
\end{align}
where $Q$ is a Gaussian random variable with mean $0$ and variance 
\begin{align}
\text{Var}(Q)= \frac{\dot{\nu}(\theta_0)^T I^{-1}(\theta_0) \dot{\nu}(\theta_0)}{2\sigma^2(\theta_0) + \dot{\nu}(\theta_0)^T I^{-1}(\theta) \dot{\nu}(\theta_0)}.
\end{align}
This concludes the proof.

\section{Details for \Cref{sec:problem_with_ppcs}} \label{appendix:problem_with_ppcs}

Suppose we observe data $\mby =\{y_i\}_{i=1}^n, y_i \sim N(\mu, \sigma^2)$ where $\sigma^2$ is known. The prior is taken to be $\mu \sim N(\mu_0, \sigma_0^2)$. 

We consider the posterior predictive $p$-value with diagnostic $d(\mby) = \overline{\mby}$. The posterior predictive distribution of $\overline{\mby}^\rep$ is:
\begin{align}
\overline{\mby}^\rep | \overline{\mby}^\obs &= \int \rmp(\overline{\mby}^\rep | \theta) \rmp(\theta | \overline{\mby}^\obs) d\theta \\
&\sim  N\left(\rho_n \overline{\mby}^\obs + (1 - \rho_n)\mu_0, \left(1 + \rho_n\right)\frac{\sigma^2}{n} \right)
\end{align}
where $\rho_n = n \sigma_0^2 / (n\sigma_0^2 + \sigma^2)$.

Then, the posterior predictive $p$-value is:
\begin{align}
\rmp(d(\mby_{rep}) > d(\yobs) | \yobs)) &= 1 - \Phi\left(\frac{\overline{\mby}^\obs - \rho_n \overline{\mby}^\obs - (1 - \rho_n)\mu_0}{\sqrt{(1 + \rho_n)\sigma^2/n}} \right).
\end{align}

\section{Discussion of the partial predictive check} \label{appendix:partial-pc-discussion}

An alternative method to obtain calibrated $p$-values is the partial predictive check of \citet{bayarri2000p}. The partial PC achieves calibration by calculating a conditional posterior predictive that is independent of the diagnostic.  However, when the diagnostic includes the sufficient statistics of the model, the partial predictive check essentially becomes a prior predictive check.  To illustrate this point, consider the partial predictive check for the test in \Cref{sec:problem_with_ppcs}. The partial predictive check uses the following predictive distribution:
\begin{align}
\rmp(\yrep | \mu) \rmp(\mu | \yobs \backslash d(\yobs)), \quad\text{where} \quad \rmp(\mu | \yobs \backslash d(\yobs))\propto \frac{\rmp(\yobs|\mu)\rmp(\mu)}{\rmp(d(\yobs)|\mu)}.
\end{align}

In this example, $\rmp(\mu | \yobs \backslash \overline{\mby}^\obs)$ is simply the prior on $\mu$, as we are removing the influence of the sufficient statistic, $\overline{\mby}^\obs$. Then, the distribution of the partial posterior predictive is
\begin{align}
\overline{\mby}^\rep| \yobs \backslash d(\yobs)\sim N\left(\mu_0, (\sigma_0^2 + \sigma^2)/n \right).
\end{align}
The partial predictive $p$-value is then:
\begin{align}
\rmp(d(\yrep) > d(\yobs) | \yobs \backslash \overline{\mby}^\obs) = 1 - \Phi\left( \frac{\overline{\mby}^\obs - \mu_0}{\sqrt{(\sigma_0^2 + \sigma^2)/n}}\right). \label{eq:partial_pc}
\end{align}
This is the prior predictive $p$-value.  That is, if the diagnostic is the only sufficient statistic, the partial predictive $p$-value coincides with the prior predictive $p$-value, which is not calibrated.  This does not contradict \citet{robins2000asymptotic,bayarri2000p}, however, who prove the partial predictive check is calibrated under certain assumptions. One of these assumptions is that the parameters of the predictive distribution converge to the MLE, which is not the case here. 

However, the partial posterior check is similar to the \gls{HPC} when the partial diagnostic is defined to use a subset of the data. In the above example, we could choose $d(\yobs)=1/(n/2)\sum_{i=1}^{n/2} y_i^{\obs} \eqqcolon \overline{\mby}^{\obs}_{1:n/2}$  In this case, the conditional posterior is:
\begin{align}
\rmp(\mu | \yobs \backslash d(\yobs)) &\propto \frac{\exp\left\{-\frac{1}{2}\sum_{i=1}^n\left[y_i^{\obs} - \mu\right]^2\right\}}{\exp\left\{-\frac{n}{4}\left[\overline{\mby}_{1:n/2} - \mu\right]^2\right\}}\rmp(\mu) \\
&\propto \prod_{i=n/2+1}^n \rmp(y^{\obs}_i | \mu)\rmp(\mu).
\end{align}
For this diagnostic, the partial posterior check is equivalent to the \gls{HPC}. 

In certain cases, the partial posterior check with a data-split diagnostic may be more efficient in its use of the data than \gls{HPC}.  This is because the \gls{HPC} will always use a subset of the data for posterior inference. A partial predictive check meanwhile may only remove a sufficient statistic of this subset of the data. Despite this, the partial predictive check can be difficult to calculate, and requires re-calculation for each diagnostic function. Meanwhile, a \gls{HPC} is simple to implement, and the inferred posterior can be used to check many different diagnostic functions.

\section{Additional details for \Cref{sec:experiments}} \label{appendix:experiments}

\begin{figure}[h]
	\begin{subfigure}[t]{0.49\textwidth}
		\centering
		\includegraphics[width = \textwidth]{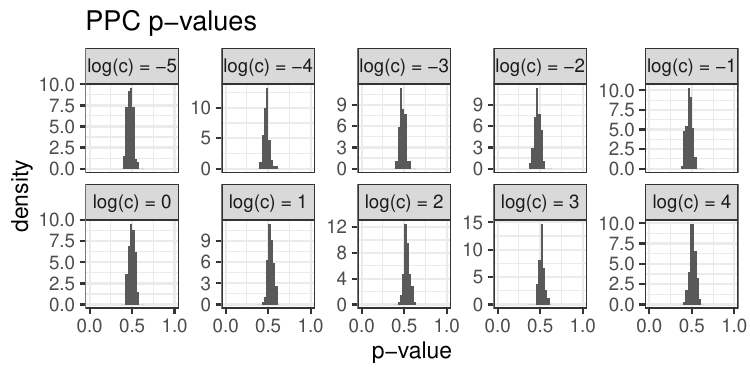}
	\end{subfigure}
	\begin{subfigure}[t]{0.49\textwidth}
		\centering
		\includegraphics[width = \textwidth]{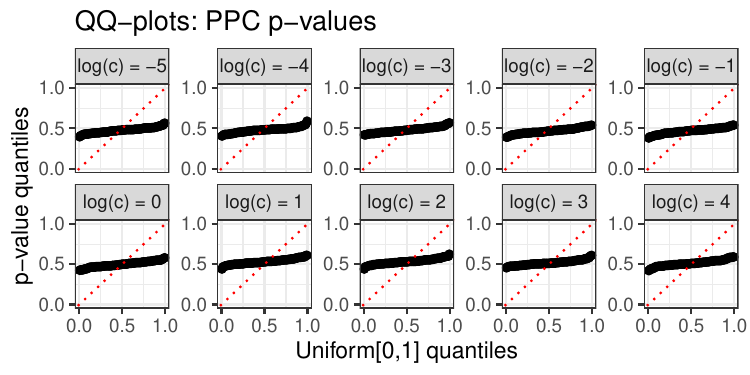}
	\end{subfigure}
	\caption{\gls{PPC} $p$-values are concentrated around 0.5 for all values of the regularization parameter $c$. Left: Histograms of \gls{PPC} $p$-values. Right: QQ-plots comparing the quantiles of the \gls{PPC} $p$-values with the quantiles of a uniform[0,1] random variable.} \label{fig:ppc_p_vals}
\end{figure}

\begin{figure}[h]
	\begin{subfigure}[t]{0.49\textwidth}
		\centering
				\includegraphics[width = \textwidth]{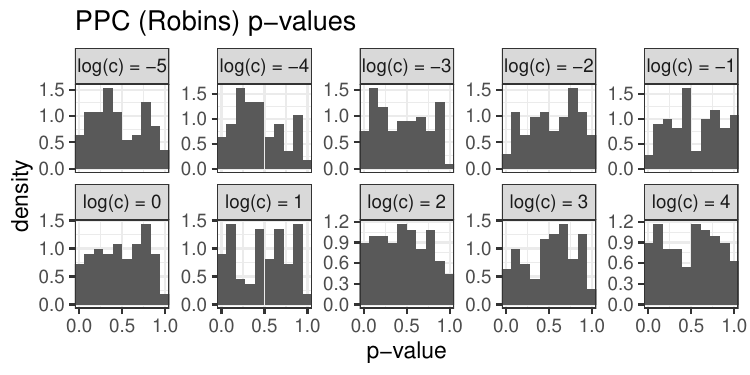}
	\end{subfigure}
	\begin{subfigure}[t]{0.49\textwidth}
		\centering
				\includegraphics[width = \textwidth]{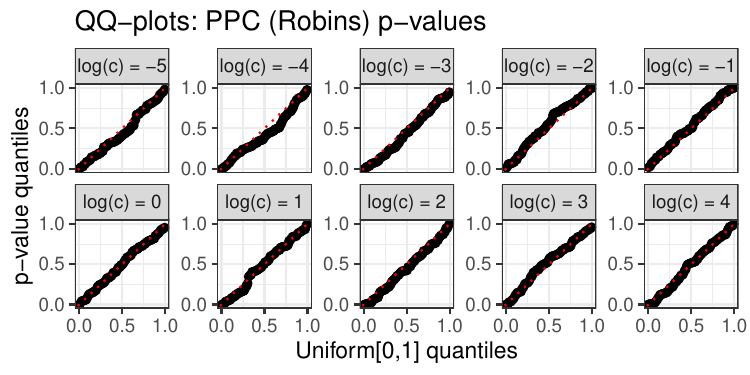}
	\end{subfigure}
	\caption{Calibrated \gls{PPC} $p$-values \citep{robins2000asymptotic} are approximately uniformly distributed but they do not detect overfitting for large values of the regularization parameter $c$. Left: Histograms of calibrated \gls{PPC} $p$-values. Right: QQ-plots comparing the quantiles of the calibrated \gls{PPC} $p$-values with the quantiles of a uniform[0,1] random variable. Plots are over $K=100$ replications of the data.}  \label{fig:ppc_calibrated_p_vals}
\end{figure}

\begin{figure}[h]
	\begin{subfigure}[t]{0.49\textwidth}
		\centering
				\includegraphics[width = \textwidth]{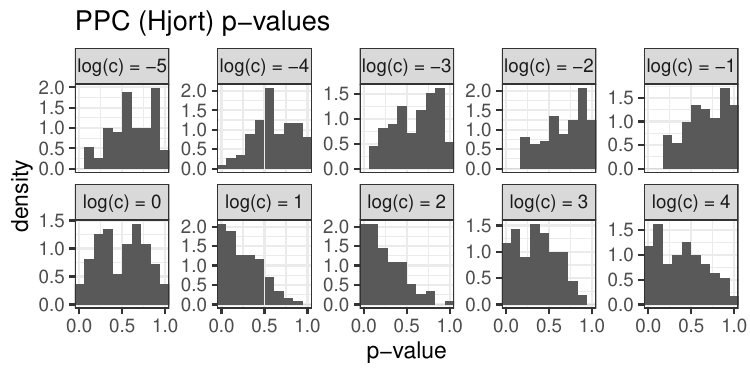}
	\end{subfigure}
	\begin{subfigure}[t]{0.49\textwidth}
		\centering
				\includegraphics[width = \textwidth]{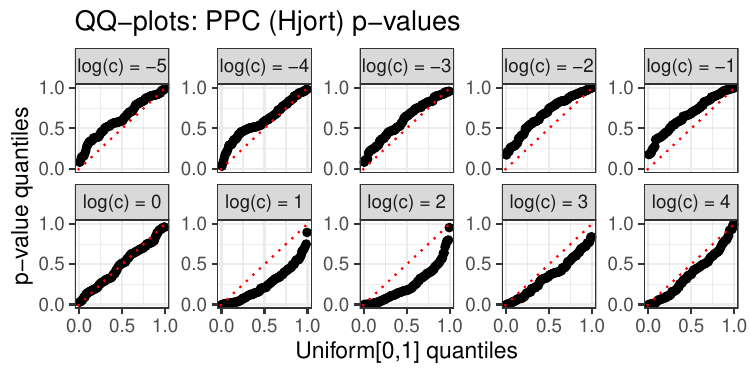}
	\end{subfigure}
	\caption{Calibrated cppp $p$-values \citep{Hjort:2006} do not detect overfitting for large values of the regularization parameter $c$. Left: Histograms of cppp $p$-values. Right: QQ-plots comparing the quantiles of the cppp $p$-values with the quantiles of a uniform[0,1] random variable. Plots are over $K=100$ replications of the data.} 
\label{fig:hjort_calibrated_p_vals}
\end{figure}

In this section, we provide additional plots for the regression study in \Cref{sec:ridge_regression}. 
\begin{itemize}
\item \Cref{fig:ppc_p_vals} shows the empirical distribution of \gls{PPC} $p$-values. The $p$-values are concentrated around 0.5 for all values of $c$.
\item \Cref{fig:ppc_calibrated_p_vals} shows the empirical distribution of \gls{PPC} $p$-values calibrated following \citet{robins2000asymptotic}. While the $p$-values are uniform, they show a similar distribution for all values of $c$ and so are unable to detect model misfit.
\item \Cref{fig:hjort_calibrated_p_vals} shows the empirical distribution of cppp $p$-values \citep{Hjort:2006}. The cppp $p$-values do not detect model misfit for large $c$.
\item \Cref{fig:pop_pc_p_vals} shows the empirical distribution of the \gls{HPC} $p$-values. For small values of $c$, the \gls{HPC} $p$-values are approximately uniform. For large values of $c$, the \gls{HPC} $p$-values concentrate around 0 (i.e. the \gls{HPC} will always reject the model when there is insufficient regularization).
\end{itemize}

